\documentclass[12pt,aip,cha,graphicx,amsmath,amssymb,reprint,onecolumn,floatfix]{revtex4-1}
\usepackage{amssymb, graphics, epsfig, color, ulem, float,graphicx,suppenv}
\usepackage{amsmath,amstext,bm,verbatim,bbold}
\usepackage{graphicx,subfigure}
\usepackage{epstopdf}
\usepackage{mathtools}
\usepackage[amssymb]{SIunits}
\usepackage{svg}
\usepackage{natbib}
\usepackage{todonotes}
\usepackage{xurl}
\newcommand{\kdet}{ k_{det}}
\newcommand{\kdetpr}{ k_{det}^\prime}
\newcommand{\koffzw}{ k_{o\!f\!f}\hspace{-0.4mm}(\sigma_{})} %
\newcommand{\koffone}{ k_{o\!f\!f}\hspace{-0.4mm}(0)}
\newcommand*{\bfrac}[2]{\genfrac{}{}{0pt}{}{#1}{#2}}

\begin{document}
\title{Twofold mechanosensitivity ensures actin cortex reinforcement upon peaks in mechanical tension}

\author{Valentin Ruffine}
\affiliation{Cluster of Excellence Physics of Life, Technische Universit\"at Dresden, Dresden, Germany}
\author{Andreas Hartmann}
\affiliation{B CUBE - Center for Molecular Bioengineering, Technische Universit\"at Dresden, Dresden, Germany}
\author{Michael Schlierf}
\affiliation{Cluster of Excellence Physics of Life, Technische Universit\"at Dresden, Dresden, Germany}
\affiliation{B CUBE - Center for Molecular Bioengineering, Technische Universit\"at Dresden, Dresden, Germany}
\affiliation{Faculty of Physics, Technische Universit\"at Dresden, Dresden, Germany}
\author{Elisabeth Fischer-Friedrich}
\email{Corresponding author: elisabeth.fischer-friedrich@tu-dresden.de}
\affiliation{Cluster of Excellence Physics of Life, Technische Universit\"at Dresden, Dresden, Germany}
\affiliation{Faculty of Physics, Technische Universit\"at Dresden, Dresden, Germany}
\affiliation{Biotechnology Center, Technische Universit\"at Dresden, Dresden, Germany}

\begin{abstract}{
The actin cortex is an active biopolymer network underneath the plasma membrane at the periphery of mammalian cells. It is a major regulator of cell shape through the generation of active cortical tension. In addition, the cortex constitutes a mechanical shield that protects the cell during mechanical agitation. Cortical mechanics is tightly controlled by the presence of actin cross-linking proteins, that dynamically bind and unbind actin filaments. Cross-linker actin bonds are weak non-covalent bonds whose bond lifetime is likely affected by mechanical tension in the actin cortex making cortical composition inherently mechanosensitive. Here, we present a quantitative study of changes in cortex composition and turnover dynamics upon short-lived peaks in active and passive mechanical tension in mitotic HeLa cells. Our findings disclose a twofold mechanical reinforcement strategy of the cortex upon tension peaks entailing i) a direct catch-bond mechanosensitivity of cross-linkers filamin and $\alpha$-actinin  and ii) an indirect cortical mechanosensitivity that triggers actin cortex reinforcement via enhanced polymerization of actin. We thereby disclose a `molecular safety belt' mechanism that protects the cortex from injury upon mechanical challenges. 
}
\end{abstract}
\maketitle

\maketitle

\section{Introduction}
Within animal cells, the protein actin assembles into a biopolymer network thereby weaving an essential material of life – the actin cytoskeleton. The actin cortex is a thin layer of interweaved actin filaments located beneath the plasma membrane of the cell. It is an active material with rich mechanical properties characterized by stress-stiffening and frequency-dependent elastic moduli \cite{broedersz2010, lieleg2010, fischer-friedrich2016, bonfanti2020}. In addition, self-generated active tension in the actin cortex gives rise to an adjustable cell surface tension that enables the dynamic regulation of cell shape \cite{salb12, kelkar2020, chalut2016}. Therefore, the actin cortex and its mechanical properties are pivotal for cellular function  \cite{salb12, kelkar2020, chalut2016}.

Within the actin cytoskeleton, actin filaments are cross-linked through transient binding of active and passive cross-linker proteins \cite{vicente-manzanares2009, murphy2015, nakamura2011}. The main active cross-linkers are nonmuscle myosin II motor proteins, and the main passive cross-linkers are $\alpha$-actinin-1, $\alpha$-actinin-4, and filamins A and B \cite{vicente-manzanares2009, murphy2015, nakamura2011}. The degree of cross-linking of the actin cytoskeleton was shown to tightly regulate cortex mechanical properties with significantly higher stiffness at larger cross-linking degrees \cite{gardel2004, koenderink2009, broedersz2010, kasza2009b}.
 It is thus not surprising that actin cross-linkers have been found to be essential for biological function; 
loss or mutation of actin cross-linkers {\it in vivo} has been reported to result in various pathological phenotypes where several of the disease-associated mutations have been shown to affect binding to f-actin  \cite{ward2008, yao2011, wade2020, sawyer2009}. 
More specifically, loss of nonmuscle myosin II was shown to be deadly to mammalian embryos \cite{conti2004, tullio1997}.
Furthermore, cell shape regulation during cell division was observed to be strongly impaired when $\alpha$-actinin-4 is knocked down \cite{mukhina2007, toyoda2017}.
Moreover, cell migration and attachment to the extracellular matrix (ECM) require filamins \cite{cunningham1992, del_valle-perez2010, liu2021, ketebo2021, xu2010, cole2021,razinia2012, nakamura2011, lynch2011} and loss or mutation of filamins leads to developmental defects \cite{sawyer2009, moutton2016, wade2020, zhou2021}.

During their lifespan, cells need to undergo a sequence of shape changes e.g. during cell division or tissue rearrangement. 
Correspondingly, the actin cortex is subject to dynamically evolving active and passive mechanical tension, where active tension is generated through cortex-associated myosin motors and passive tension is induced through cell deformation via cell-internal or external forces.  
At the level of individual cortical cross-linkers, both tensions will induce tensile forces on the bonds between cross-linkers and f-actin \cite{broedersz2008,kasza2009,mulla2018,mulla2022}. 
In such transient non-covalent bonds between proteins, tensile force is usually expected to increase the unbinding rate \cite{bell1978}. 
Thus, tensile forces in the cortex could easily induce a chain reaction of rupturing cross-linker bonds which could lead to material failure and cellular injury \cite{mulla2018}. 
However, previous studies have suggested that the bonds between some cross-linkers and f-actin are catch bonds, namely that their lifetime increases with increasing tensile force. In particular, individual nonmuscle myosin II molecules have been shown to bind longer to f-actin when forces that oppose their motor work are increased \cite{kovacs2007, norstrom2010}. 
Additionally, experimental studies in live cells and with single-molecule experiments {\it in vitro} confirmed catch-binding kinetics for $\alpha$-actinin-4 \cite{hosseini2020, mulla2022}. 
Catch-binding of cross-linkers provides the actin cortex with a mechanosensitive way to reinforce itself mechanically through an increased cross-linking density when it is under conditions of high mechanical stress \cite{mulla2018, mulla2022}.
On the other hand, mechanical stimuli were shown to be able to trigger indirect mechanosensitive responses of the actin cytoskeleton via biochemical signaling cascades (e.g. mediated by Calcium influx) leading to a change in actin polymerization dynamics  \cite{miroshnikova2021, wan2013, kaplan2022, gavara2008, glogauer1998}.

In this study, we present a comprehensive investigation of cortical composition and dynamics upon induced changes of active and passive deformation-induced tension. 
As a biological model system, we used transgenic mitotic HeLa cells expressing fluorescently labeled cortical proteins of interest. 
Tension changes were either imposed i) through cell squeezing with the cantilever of an atomic force microscope inducing passive mechanical tension peaks or ii) through chemical manipulation of myosin motor activity inducing peaks in active tension. 
We focused on the mechanosensitive response of the scaffold proteins f-actin and myosin II as well as of the cross-linker proteins $\alpha$-actinin-1 and filamins A and B. 
Our findings disclose that the cortex exhibits a twofold mechanosensitive response that includes i) a short-term direct mechanosensitivity of the actin cortex mediated by a catch bond behavior of passive cross-linker proteins and 
ii) an indirect cortical mechanosensitivity that triggers long-term cortex reinforcement through increased actin polymerisation in particular after active tension peaks.
Our findings provide an explanation as to how the actin cortex can stay structurally intact in the presence of high tension while maintaining fast molecular turnover and molecular adaptability.

\section{Results}
We first asked the question, to which extent the degree of cross-linking of the actin cortex depends on the mechanical tension in the cortical network. If bonds between cross-linker molecules and f-actin were mechanosensitive, tensile forces on cross-linker bonds should change the bond lifetime and thereby the overall concentration of cross-linking molecules in the cortex. In particular, catch-binding would increase the degree of cross-linking and enrich cross-linkers at the cortex in response to increased tension \cite{mulla2022}. 
To rationalize the dependence of cross-linker binding dynamics on cortical tension, we utilized a previously presented kinetic binding model\cite{hosseini2020}, see Fig.~\ref{fig:FRAP}a. 
The model distinguishes three types of cross-linker populations --- a cytoplasmic population with a spatially homogeneous concentration $c_{cyt}$, and two cortex-associated populations where one is bound via a single binding site and the second is bound via two binding sites and thus cross-linking. They are described by the uniform surface concentration fields $c_{sb}$ and $c_{cl}$, respectively.
The parameters of the model quantify the rates of binding and unbinding as illustrated in Fig.~\ref{fig:FRAP}a. Furthermore,  the mechanosensitivity parameter $\alpha$ sets the relationship between the unbinding rate $k_{off}$ and the cortical tension, see Eq.~\eqref{eq:TaylorExpRate} and \cite{hosseini2020}. In this way, the model can predict concentration changes of cortical cross-linkers in dependence of the degree of their bond mechanosensitivity. 

For our experimental work, we used human cells (HeLa cells) which were genetically modified to express one cross-linker species with a fluorescent tag. In this way, fluorescence intensity in confocal microscopy images could be used as a readout of cross-linker concentration inside the cell in dependence of cellular localization. We chose to work with cells in mitosis as those are naturally in a round, weakly adherent state which provides a well-defined reference geometry. To avoid perturbation through cell cycle progression, cells were arrested in mitosis via co-incubation with S-trityl-L-cysteine ($2\,\mu$M), see Materials and Methods. 
In mitotic cells, the actin cortex forms a quasi continuous shell under the plasma membrane at which  cortex-associated cross-linkers are enriched. 
Therefore, in confocal images of the equatorial cross-section of cells expressing fluorescently-labeled cross-linker proteins, the cortex appears as a fluorescent halo and can be clearly distinguished from the cytoplasm, see Fig.~\ref{fig:FRAP}a,c and d and Fig.~\ref{fig:S_FRAP_img}a-c.

\subsection{Binding dynamics of actin cross-linkers depends on cortical tension in steady state}
\begin{figure}[ht]
\centering
\includegraphics[width=0.9\linewidth]{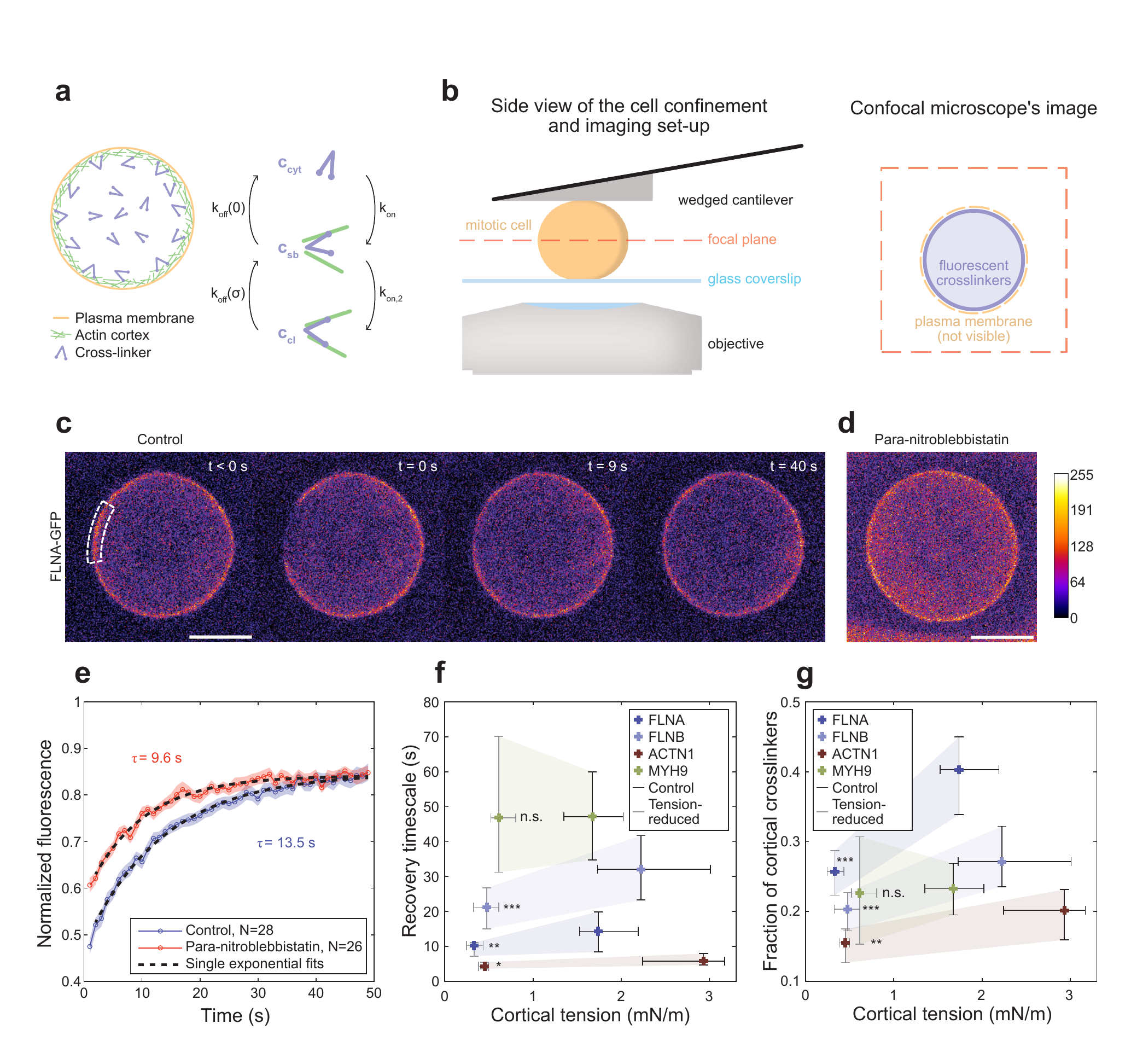}
\caption{\label{fig:FRAP}
Cortical residence time and cortical localization of actin cross-linkers change upon pharmacologically induced cortical tension reduction. 
(a) Model of cross-linker binding dynamics at different values of cortical tension $\sigma$, see \cite{hosseini2020}. 
(b) Left panel: AFM confinement assay combined with confocal microscopy applied to mitotic cells.
Right panel: schematic of a confocal image from the cell's equatorial cross-section. 
(c) Exemplary fluorescence recovery after photobleaching (FRAP) in a HeLa cell expressing fluorescently labeled filamin A (FLNA-GFP) in the control condition. The dashed white line indicates the approximate boundaries of the region bleached at $t=0~\mathrm{s}$.
(d) Exemplary confocal image of fluorescently labeled filamin A  after treatment with $10~\mathrm{\mu M}$ para-nitroblebbistatin (before bleaching). (c,d) Scale bars: $10~\mathrm{\mu m}$. The range of pixel values was adjusted in the same way in (c) and (d) and color scale of pixel values is common to both panels.
(e) Averaged FRAP curves for cells expressing fluorescently labeled filamin A (blue: control, red: tension-reduced via treatment with $10~\mathrm{\mu M}$ para-nitroblebbistatin). 
Shaded areas in either color represent the standard error of the mean. 
Dashed black lines show fits of a mono-exponential recovery, see Materials and Methods. 
Respective fit timescales are indicated in corresponding colors. 
(f,g) Recovery time scale and cortical association in control and tension-reduced condition for actin cross-linkers filamin A (FLNA-GFP; control, N=28; tension-reduced, N=26), filamin B (FLNB-GFP; control, N=25; tension-reduced, N=29), $\alpha$-actinin-1 (ACTN1-mOrange2; control, N=23; tension-reduced, N=19) and myosin II (MYH9-mKate2; control, N=29; tension-reduced, N=21). 
Tension reduction was achieved by co-incubation with either $10~\mathrm{\mu M}$ para-nitroblebbistatin (filamins and $\alpha$-actinin-1) or $2.5~\mathrm{\mu M}$ blebbistatin (myosin II).
Thick crosses indicate median values, and whiskers show the $25^\mathrm{th}$ and $75^\mathrm{th}$ percentiles. Shaded areas are guides to the eye. 
(f) Timescales were obtained from mono-exponential fit of FRAP recovery curves of individual cells, see Materials and Methods.
(g) Cortical association was quantified as the estimated fraction of cortex-associated fluorescence, see Materials and Methods and {} \cite{hosseini2020}. 
All tension differences upon treatment are statistically significant with $p<10^{-3}$. P-values were calculated using the Wilcoxon rank sum test.
}
\end{figure}
In order to probe cross-linker binding in dependence of cortical tension, we first performed experiments where Fluorescence Recovery After Photobleaching (FRAP) was measured through confocal time-lapse imaging.  Experiments were performed in conditions of high and low cortical tension and cortical tension was monitored concomitantly throughout the experiment using our previously described AFM confinement setup. In this setup, cells are confined between an AFM cantilever and the glass bottom dish and the AFM force readout is used to determine cortical tension as previously described \cite{hosseini2020}, see Fig.~\ref{fig:FRAP}b. We note that FRAP experiments were carried out in a condition of steady state of cortical tension, where viscoelastic stresses due to cantilever cell confinement had already been dissipated due to an adequate waiting time ($\ge 60\,$s) between application of cell confinement and FRAP experiments, leaving only active tension contributions in the cell cortex during the measurement \cite{fischer-friedrich2016}.
For FRAP measurements, confined HeLa cells expressing fluorescently-labeled filamin A, filamin B, $\alpha$-actinin-1 or myosin II were imaged in their equatorial cross-section with a confocal microscope, see Materials and Methods and Fig.~\ref{fig:FRAP}c,d and Fig.~\ref{fig:S_FRAP_img}a-c. 
After initial imaging of a cell's cross-section, we photobleached a small region of the actin cortex, see Fig.~\ref{fig:FRAP}c. Subsequently, we recorded a time series of confocal images and extracted fluorescence recovery curves from the recorded fluorescence signal, see Fig.~\ref{fig:FRAP}c,e.
Cell-averaged recovery curves for fluorescently labeled filamin A are shown in Fig.~\ref{fig:FRAP}e in conditions of high tension (untreated cells, blue curve) and tension-reduced conditions (red curve). Tension reduction was achieved through pharmacological inhibition of actin-associated molecular motor proteins myosin II via  co-incubation with the chemical para-nitroblebbistatin ($10 \,\mu$M), see Materials and Methods. Dashed black lines show a mono-exponential fit of the respective filamin A recovery curve  with fitted recovery time scales $\tau=13.5\,$s (control) and $\tau=9.6\,$s (tension-reduced), pointing to a faster recovery in tension-reduced conditions. 
Fitting recovery curves of individual cells for different actin cross-linkers, we found recovery time scales that are significantly reduced at tension-reduced conditions  for all cross-linkers but myosin II, see Fig.~\ref{fig:FRAP}f. 
Furthermore, we estimated the fraction of cortex-associated proteins from images before bleaching for all cross-linkers under consideration as described before \cite{hosseini2020}. We find that cortical association is significantly lowered through tension reduction  for all cross-linkers but myosin II, see Fig.~\ref{fig:FRAP}g. 
We conclude that all cross-linkers but myosin II exhibit hallmarks of catch-binding dynamics in FRAP experiments. 

\subsection{Cross-linkers increase transiently at the cortex upon passive deformation-induced tension peaks}
\begin{figure}[ht]
\centering
\includegraphics[width=0.9\linewidth]{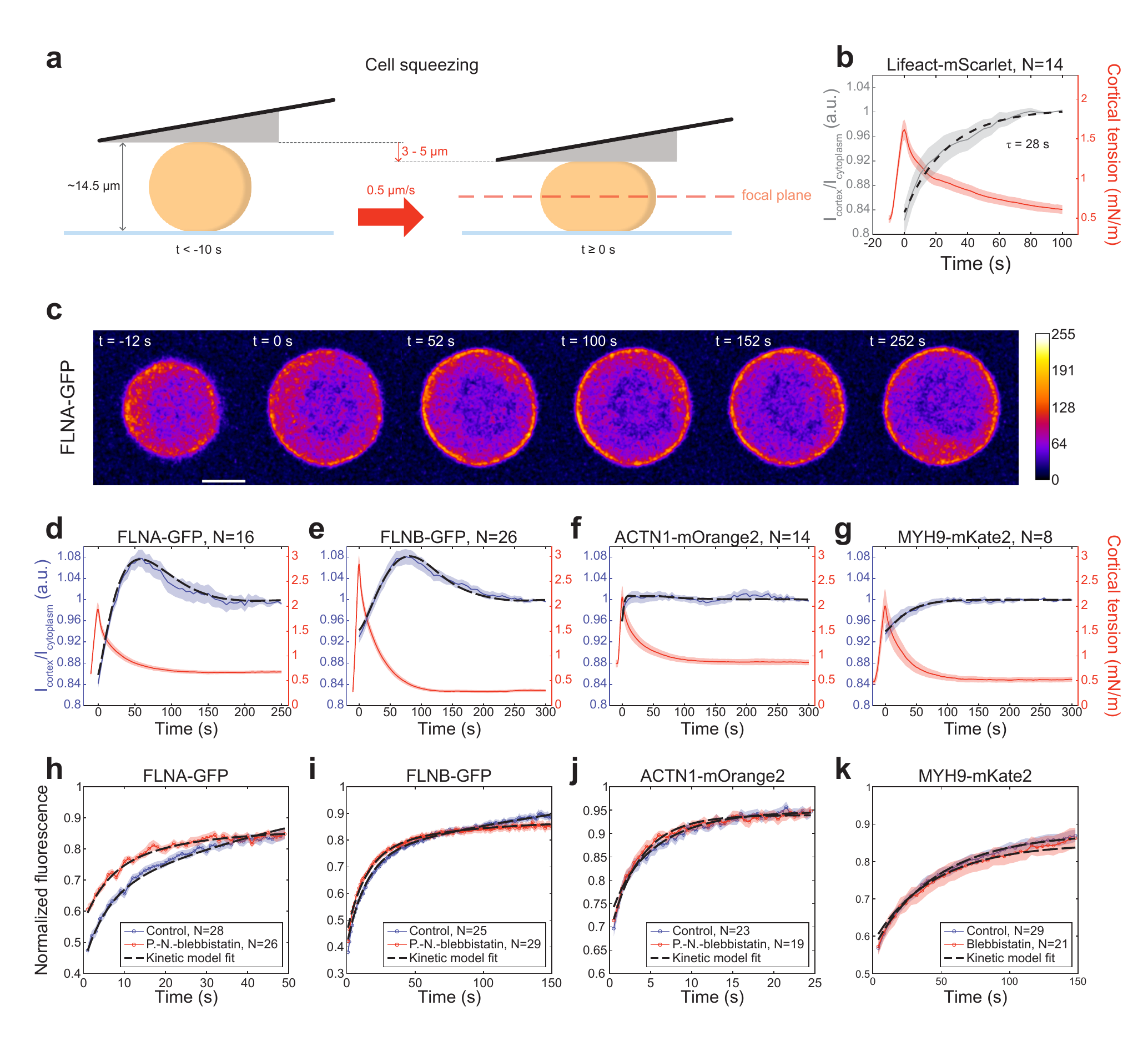}
\caption{\label{fig:Squish}
Cross-linkers respond to deformation-induced tension peaks in a way that is consistent with FRAP data.
(a) Schematic of cell-squeezing assay with the aid of an AFM cantilever (grey triangle: wedge of AFM cantilever, orange: cell, blue: dish bottom).
(b) Time evolution of cortical tension and ratio of cortical to cytoplasmic fluorescence obtained in response to AFM squeezing of mitotic HeLa cells expressing fluorescently labeled Lifeact (marker of f-actin). Solid curves (cortex-to-cytoplasm ratio: grey, cortical tension: red) show the mean value and shaded areas indicate the standard error of the mean. The dashed black line shows the fit of a mono-exponential recovery.
(c) Time series of confocal images of the equatorial cross-section of a confined mitotic HeLa cell expressing fluorescently labeled filamin A during a cell-squeezing experiment. Time $t=-12\,$s corresponds to the time point immediately before squeezing started, and time $t=0\,$s to the time point when the cantilever reached its final position and squeezing stopped.
(d-g) Time evolution of cortical tension and ratio of cortical to cytoplasmic fluorescence obtained during AFM squeezing of mitotic HeLa cells. Cells were expressing diverse fluorescently labeled cross-linkers: (d) filamin A, (e) filamin B, (f) $\alpha$-actinin-1, (g) myosin II. 
Solid curves (cortex-to-cytoplasm ratio: blue, cortical tension: red) show the mean value and shaded areas indicate the standard error of the mean. 
Dashed black lines show model fits where FRAP and squeezing data were jointly fitted, see Materials and Methods.
(h-k) Averaged FRAP curves for cells expressing fluorescently labeled filamin A (data as in Fig.~\ref{fig:FRAP}e), filamin B, $\alpha$-actinin-1 or myosin II
(solid blue line: control, solid red line: tension-reduced via treatment with $10~\mathrm{\mu M}$ para-nitroblebbistatin or $2.5~\mathrm{\mu M}$ blebbistatin). 
Shaded areas in either color represent the standard error of the mean. 
Dashed black lines show fits of the model, see Materials and Methods.
}
\end{figure}

So far, we examined the influence of cortical tension on cortical cross-linker concentrations in steady state conditions without tension dynamics. 
To investigate how cross-linkers respond to dynamic tension changes, we perturbed confined cells in steady state by a rapid deformation through a decrease of confinement height, see Fig.~\ref{fig:Squish}a. This cell-squeezing was previously reported to be associated with a dilation of the cortex and a corresponding transient peak in cortical tension, see red curves in Fig.~\ref{fig:Squish}b and d-g, and {} \cite{mokbel2020,fischer-friedrich2016}.
In conjunction, we monitored the ratio of cortical to cytoplasmic fluorescence of actin cross-linkers and f-actin. This ratio was derived from confocal images of equatorial cell cross sections during dynamic tension evolution, see Fig.~\ref{fig:Squish}b-g. 
For imaging of f-actin, we used fluorescently tagged Lifeact --- an established label of f-actin  \cite{riedl2008}.

We observed that, after cell squeezing comes to a halt, f-actin concentration increases at the cortex and relaxes to a new equilibrium value with a relaxation time scale of $\tau\approx 28\,$s, see black dashed line in Fig.~\ref{fig:Squish}b. This observation is consistent with the idea of a preceding dilution of cortical f-actin induced by squeezing-associated cell surface area dilation. Once the AFM cantilever and surface area dilation come to a halt (time $t=0\,$s), actin polymerization kinetics evolves the diluted f-actin concentration over time to a steady state concentration which is higher than the initial concentration. However, we note that this increase in f-actin might also point to a second kind of cortical mechanosensitivity that comes on top of cross-linker catch-binding. 
 
For actin cross-linkers with a sufficiently strong catch-binding, we expect a peak of cortical intensity right after the peak in cortical tension as a hallmark of tension-induced reduction in cross-linker unbinding. This effect should be visible in addition to the relaxation to a new steady state concentration in response to changes of the cortical substrate f-actin. 
Observing cortex-to-cytoplasm ratios of actin cross-linkers filamin A, filamin B, $\alpha$-actinin-1 and myosin II, we see this mechanosensitive hallmark in the dynamics of cortex-to-cytoplasm ratios of filamin A and filamin B, which show a peak at $\approx 60\,$s and $80\,$s after the tension peak, respectively, see Fig.~\ref{fig:Squish}c-e. 
By contrast, $\alpha$-actinin-1 and myosin II show no corresponding peak in cortical intensity in response to a preceding tension peak, see Fig.~\ref{fig:Squish}f and g. 
We conclude that our study of deformation-induced tension peaks gives evidence that the actin-binding of filamins A and B is mechanosensitive. On the other hand, we find no evidence for mechanosensitivity of $\alpha$-actinin-1 and myosin II in this assay.

\subsection{Model fitting discloses diverse mechanosensitivity of actin cross-linker binding}
Next, we simultaneously fitted our kinetic binding model to the results of our FRAP and cell deformation experiments.
For that purpose, we performed  weighted least square fits of FRAP recoveries and deformation-induced changes of tension and cortex-to-cytoplasm ratios, see Materials and Methods. 
As part of this scheme, analytical solutions of cortical FRAP recovery (see Eq.~\eqref{eq:FRAP}) were fitted to averaged experimental FRAP recovery curves in control (high tension) and tension-reduced conditions. Furthermore, numerical simulations of deformation-induced tension responses were fitted to  measured data, see the Materials and Methods. 
In particular, this simulation took into account the increase of f-actin following imposed cell deformation, see Fig.~\ref{fig:Squish}b. As f-actin is the substrate for cross-linker binding, a proportional change in cross-linker binding rate was assumed, see Materials and Methods.

This method yielded model parameters of kinetic binding for $\alpha$-actinin-1, filamins A and B and myosin II, see Table~\ref{tab:STab1}. Corresponding fits show a good agreement with the data, see black dashed curves in Fig.~\ref{fig:Squish}d-g and h-k for the time evolution of the cortex-to-cytoplasm ratio upon cell deformation and for the FRAP recoveries, respectively. 
Additionally, we used this new fitting method to {\it de novo} analyze the experimental data for $\alpha$-actinin-4 published by Hosseini \textit{et al.} \cite{hosseini2020} now including the f-actin recovery after cell deformation. 
This yields similar model parameters but, in particular, a slightly larger mechanosensitivity value $\alpha$, see Table~\ref{tab:STab1}.

\begin{figure}[pht]
\centering
\includegraphics[width=0.9\linewidth]{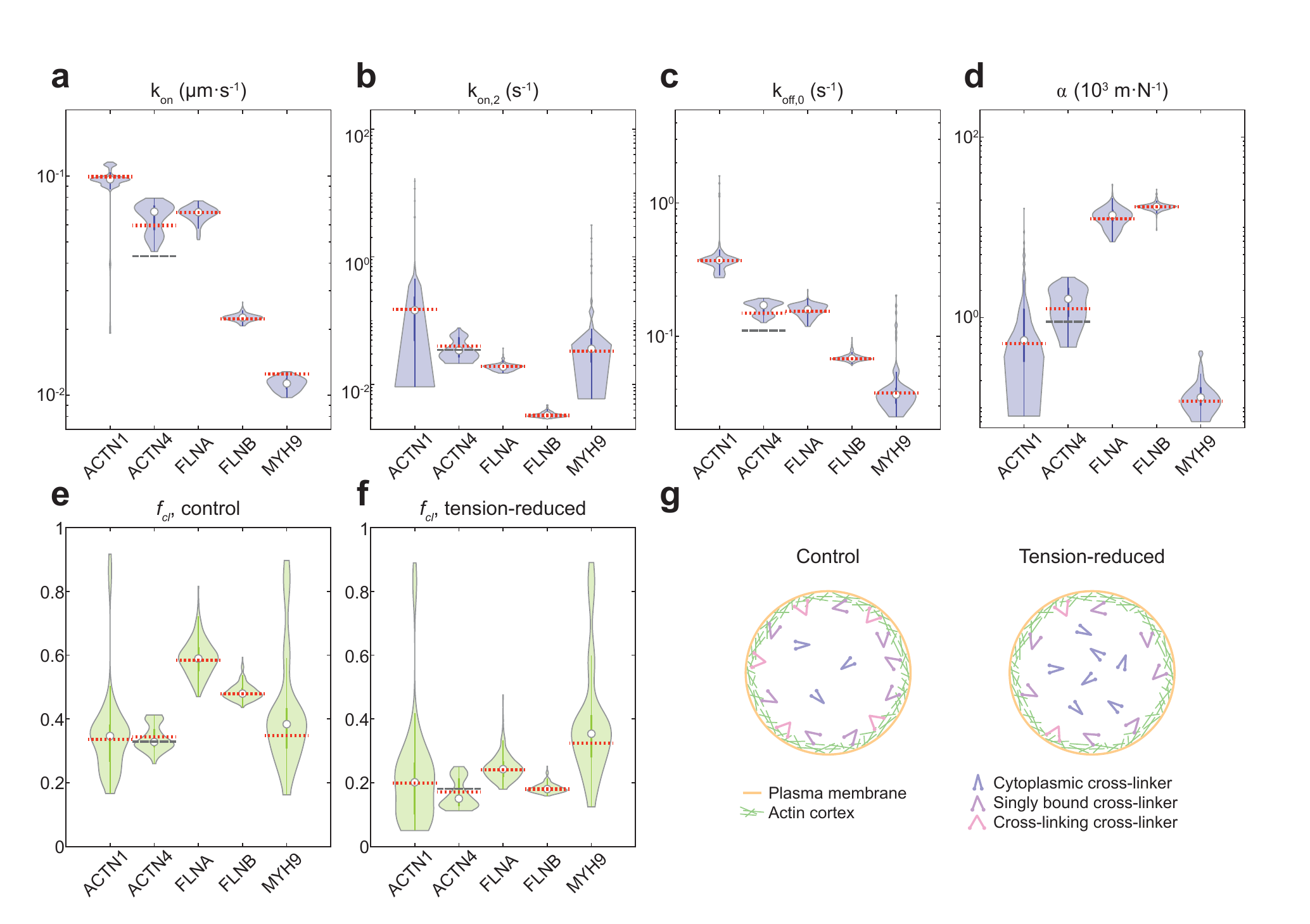}
\caption{\label{fig:ViolinPlots}
Model parameters of cross-linker binding reaction kinetics. 
(a-d) Distributions of model parameters obtained by bootstrapping and model fitting applied to FRAP and cell-squeezing data (see Materials and Methods). The model was fitted to data measured on cells expressing either $\alpha$-actinin-1, $\alpha$-actinin-4 (data from Hosseini \textit{et al.}\cite{hosseini2020}), filamin A, filamin B, or myosin II. Red dashed lines indicate the parameter values obtained by simple fitting of the average of the full dataset, i.e. without bootstrapping. Grey dashed lines indicate previously reported parameters for $\alpha$-actinin-4 {} \cite{hosseini2020}.
(e-f) Distributions of fractions $f_{cl}=N_{cl}/(N_{cl}+N_{sb})$ of cross-linking molecules at the cortex in control and tension-reduced  conditions, see Eq.~\eqref{eq:fcl}. 
Fractions were calculated using parameter distributions of (a-d) and median values of the cortical tensions measured during FRAP experiments (see Fig.\ref{fig:FRAP} and Hosseini \textit{et al.}\cite{hosseini2020}).
(g) Schematic of the distribution of cross-linkers among the three states of our model, (left) in the control condition and (right) in the tension-reduced condition, for catch-binding cross-linkers. In the tension-reduced condition, fewer cross-linkers localize at the cortex, and fewer cortical cross-linkers actually cross-link actin filaments.
}
\end{figure}

To assess the error on the fitted parameters, we used a bootstrapping analysis fitting subsets of experimental data, see Materials and Methods. Corresponding distributions of binding parameters for all actin cross-linkers are shown in  Fig.~\ref{fig:ViolinPlots}a-d. Parameter values obtained for averages of the complete dataset (see Table~\ref{tab:STab1}) are shown by red dashed lines. Moreover, Table \ref{tab:Tab1} summarizes the results of the bootstrapping analysis.

Our results show that actin cross-linkers filamins A and B as well as $\alpha$-actinin-4 exhibit a  mechanosensitivity parameter $\alpha$ significantly larger than zero. However, the mechanosensitivity of filamins ($\alpha=13.6\cdot 10^3$m/N and $\alpha=17.0\cdot 10^3$m/N, respectively) is approximately one order of magnitude bigger than that of $\alpha$-actinin-4 ($\alpha=1.61\cdot 10^3$m/N). 
This suggests that all three proteins bind actin as catch binders but with a stronger mechanosensitivity in filamins A and B as compared to $\alpha$-actinin-4. 
By contrast, the mechanosensitivity parameter fitted on $\alpha$-actinin-1 is significantly smaller ($\alpha=0.56\cdot 10^3$m/N) reflecting that effects of mechanosensitivity are less obvious in FRAP recoveries and cell deformation responses. 
Furthermore, the mechanosensitivity parameter for myosin II is  $\alpha=0.13\cdot 10^3$m/N and therefore more than one order of magnitude smaller than those for filamins and $\alpha$-actinin-4. This value of $\alpha$ corresponds to only negligible changes of binding kinetics in the explored tension range of $0-2\cdot 10^{-3}$~N/m, see Fig.~\ref{fig:FRAP}f.

Fitted model parameters enable the calculation of the estimated concentration of cross-linkers in each of the three binding states as a function of the cortical tension. In particular, this allows to infer the fraction of cross-linking molecules at the cortex, for each of the studied cross-linkers, see Eq.~\eqref{eq:fcl} in Materials and Methods, Tables \ref{tab:Tab1},~\ref{tab:STab1}, and Fig.~\ref{fig:ViolinPlots}e-g.
For filamins, we find that theory predicts an increase in the fraction of cross-linking molecules at the cortex by a factor of $\approx 2.5$ when going from a tension-reduced state to a state of high cortical tension (control condition for mitotic cells). 
For $\alpha$-actinin-4 and $\alpha$-actinin-1, by comparison, this increase is approximately twofold.
We note that for $\alpha$-actinin-1 predicted changes have larger error margins, see Table \ref{tab:Tab1}. 

For myosin II, the predicted change of the fraction of cross-linking proteins at the cortex is negligible in the light of fitting errors, see Table~\ref{tab:Tab1}, columns 10 and 12. Correspondingly, our data suggest that cortical lifetimes of myosin II do not show a significant mechanosensitivity in the explored tension range. 
(While we are aware that the molecular details of myosin II binding to the actin cortex are not well represented by our simple model, we assume that obtained model parameters are still able to report hallmarks of mechanosensitivity in actin binding of myosin II filaments if they are present.)

\begin{table}[ht]
\centering
\begin{tabular}{|l||c|c|c|c|c|c|c|c|c|c|c|c|}
    \cline{2-13}
     \multicolumn{1}{c|}{}& $k_{on}$ & IQR & $k_{on,2}$ & IQR & $\koffone$ & IQR & $\alpha$ & IQR & $f_{cl}$ & IQR & $f_{cl}$ & IQR \\
     \multicolumn{1}{c|}{}& $\mathrm{(\mu m/s)}$ & & $\mathrm{(1/s)}$ & & $\mathrm{(1/s)}$ & & $\mathrm{(10^3m/N)}$ & & Ctrl & & tension-reduced & \\
    \hline\cline{2-13}
     ACTN1 & 9.65e$-$2 & $\bfrac{9.38\mathrm{e}-2}{9.87\mathrm{e}-2}$ & 1.47e$-$1 & $\bfrac{4.83\mathrm{e}-2}{2.37\mathrm{e}-1}$ & 3.72e$-$1 & $\bfrac{3.48\mathrm{e}-1}{3.91\mathrm{e}-1}$ & 5.60e$-$1 & $\bfrac{3.25\mathrm{e}-1}{1.24}$ & 35~\% & $\bfrac{27~\%}{38~\%}$ & 20~\% & $\bfrac{10~\%}{26~\%}$\\
     \hline
     ACTN4 & 6.87e$-$2 & $\bfrac{5.66\mathrm{e}-2}{7.32\mathrm{e}-2}$ & 3.44e$-$2 & $\bfrac{2.68\mathrm{e}-2}{5.46\mathrm{e}-2}$ & 1.71e$-$1 & $\bfrac{1.45\mathrm{e}-1}{1.79\mathrm{e}-1}$ & 1.61 & $\bfrac{1.02}{2.12}$ & 33~\% & $\bfrac{31~\%}{37~\%}$ & 15~\% & $\bfrac{13~\%}{21~\%}$\\
     \hline
     FLNA & 6.84e$-$2 & $\bfrac{6.53\mathrm{e}-2}{7.15\mathrm{e}-2}$ & 1.90e$-$2 & $\bfrac{1.75\mathrm{e}-2}{2.00\mathrm{e}-2}$ & 1.58e$-$1 & $\bfrac{1.48\mathrm{e}-1}{1.71\mathrm{e}-1}$ & 13.6 & $\bfrac{11.7}{15.6}$ & 59~\% & $\bfrac{55~\%}{62~\%}$ & 24~\% & $\bfrac{22~\%}{27~\%}$\\
     \hline
     FLNB & 2.26e$-$2 & $\bfrac{2.21\mathrm{e}-2}{2.30\mathrm{e}-2}$ & 3.30e$-$3 & $\bfrac{3.16\mathrm{e}-3}{3.54\mathrm{e}-3}$ & 6.81e$-$2 & $\bfrac{6.67\mathrm{e}-2}{7.06\mathrm{e}-2}$ & 17.0 & $\bfrac{16.2}{17.6}$ & 48~\% & $\bfrac{47~\%}{50~\%}$ & 18~\% & $\bfrac{17~\%}{19~\%}$\\
     \hline
     MYH9 & 1.13e$-$2 & $\bfrac{1.07\mathrm{e}-2}{1.19\mathrm{e}-2}$ & 3.64e$-$2 & $\bfrac{2.23\mathrm{e}-2}{5.21\mathrm{e}-2}$ & 3.64e$-$2 & $\bfrac{3.12\mathrm{e}-2}{4.07\mathrm{e}-2}$ & 1.30e$-$1 & $\bfrac{1.06\mathrm{e}-1}{1.68\mathrm{e}-1}$ & 38~\% & $\bfrac{31~\%}{43~\%}$ & 35~\% & $\bfrac{28~\%}{41~\%}$\\
     \hline
\end{tabular}
\caption{Medians and interquartile ranges (IQR) of the distributions of model parameters $k_{on}, \, k_{on,2}, \, \koffone$ and $\alpha$ obtained by bootstrapping as shown in Fig.~\ref{fig:ViolinPlots}.
Further, the medians and IQRs of the fractions $f_{cl}=N_{cl}/(N_{cl}+N_{sb})$ (see Eq.~\eqref{eq:fcl}) of cross-linking molecules at the cortex are given for control and tension-reduced conditions. 
Fractions were calculated using the model parameter distributions and median values of the cortical tensions measured during FRAP experiments, see Fig.\ref{fig:FRAP}.
For $\alpha$-actinin-4, these tensions were measured by Hosseini {\it et al.} and were $2.323$ and $0.433~\mathrm{mN/m}$, respectively \cite{hosseini2020}.}
\label{tab:Tab1}
\end{table}

\subsection{Tension-sensitive FLIM-FRET confirms model predictions of cross-linking filamin A fractions}
\begin{figure}[ht]
\centering
\includegraphics[width=0.9\linewidth]{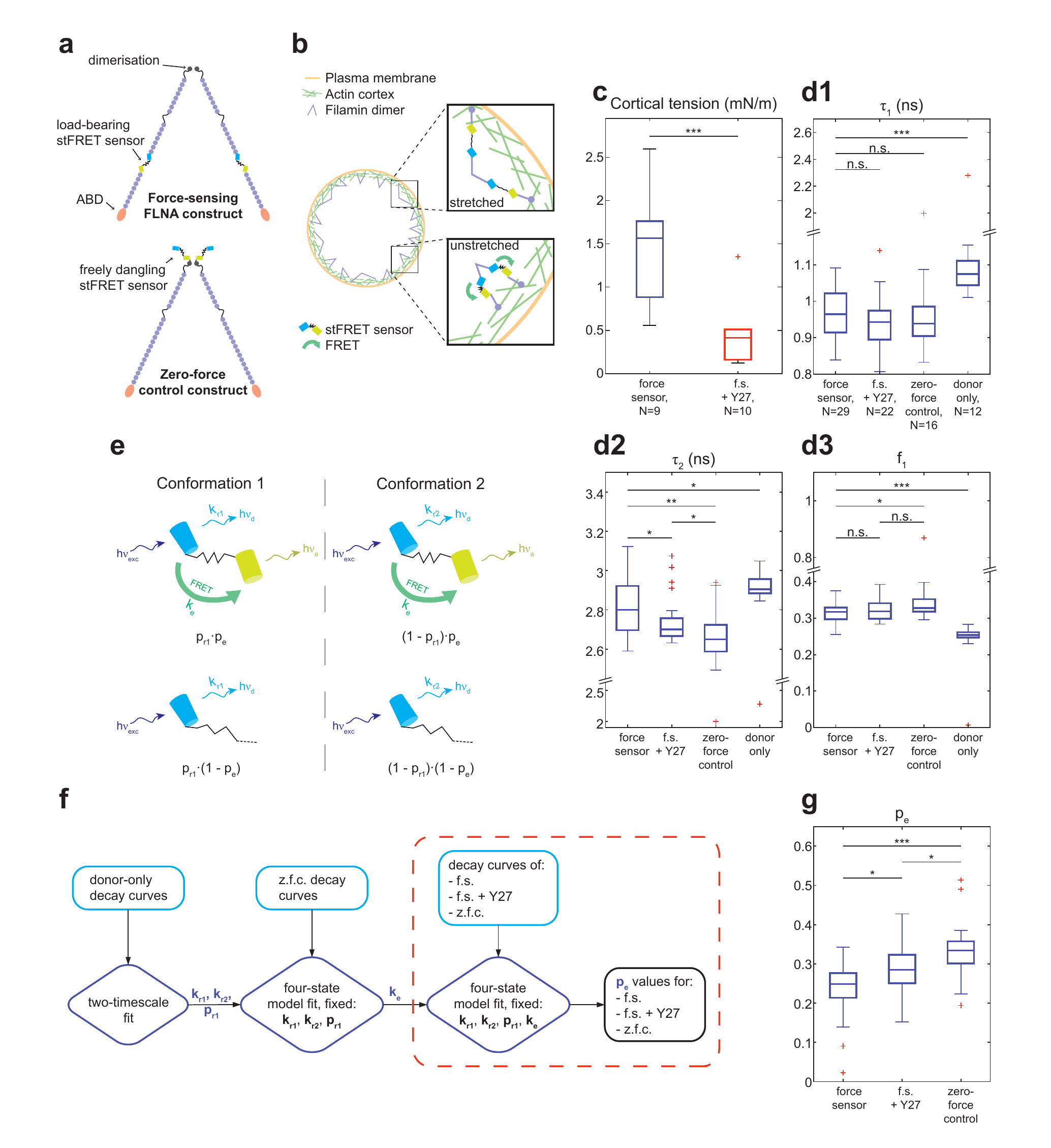}
\caption{\label{fig:FRET}
FLIM-FRET measurements confirm that the fraction of stretched filamin A cross-linkers at the cortex increases with cortical tension.
(a) Schematics of the force-sensing (f.s., top) and zero-force control (z.f.c., bottom) filamin A constructs with the stFRET sensor placed in different positions \cite{meng2008}. 
The stFRET sensor consists of a donor and an acceptor fluorophores connected by an $\alpha$-helix linker. 
(b) Schematic of force-sensing filamin A stFRET constructs in the cortex of a mitotic cell. Filamins that are cross-linking can get stretched such that the distance between acceptor and donor exceeds the range of detectable FRET.
(c) Cortical tension of mitotic HeLa cells expressing the force-sensing FRET sensor in control and tension-reduced conditions (co-incubation with Y-27632, $5~\mu$M).
(d) Results of two-exponential fitting of fluorescence lifetime distributions of the FRET donor signal measured on cells expressing either the force-sensing construct (control condition: force sensor, tension-reduced condition: f.s. + Y27), the zero-force control construct, or mCerulean (donor only).
(d1) Shortest timescale, (d2) longest timescale, and (d3) fraction of photons emitted at the shortest timescale. 
(e) Schematic illustrating possible emission states in the four-state fluorescence lifetime model in the stFRET sensor. 
Donors can be in one of two conformations, and either be able to undergo FRET (top row) or not (bottom row). The state probabilities are indicated under the corresponding state schematics. (Rate of FRET: $k_e$, rates of radiative decay: $k_{r,1}$ and $k_{r,2}$, probability to be in a FRET-enabled state: $p_e$, probability to be in conformation 1: $p_{r,1}$).
(f) Representation of the workflow used for analysis of the FLIM-FRET data measured on cells expressing either the force-sensing construct (control: f.s., tension-reduced: f.s. + Y27), the zero-force control construct (z.f.c.), or mCerulean (donor only).
(g) Fitted probability $p_e$ for cells expressing the force-sensing FRET construct with and without tension-reducing drug treatment (left and middle boxplot) and cells expressing the zero-force control FRET construct (right boxplot). P-values are calculated using the Wilcoxon rank sum test.
}
\end{figure}
In order to further substantiate our findings of cross-linker catch-binding, we employed fluorescence lifetime imaging of F\"orster Resonance Energy Transfer (FLIM-FRET) to quantify the fraction of cross-linking filamin A molecules in the cortical area and to provide benchmarking of measured binding kinetics. 
To this end, we investigated changes in cross-linker stretching in dependence of cortical tension using the FRET-based force sensor stFRET inside filamin A, see Fig.~\ref{fig:FRET}a, Materials and Methods and {} \cite{meng2008}. When cortex-associated filamin A is in a cross-linking state, cortical tension translates into a tensile force in the cross-linker that can stretch the FRET pair linker inside the stFRET sensor and thereby reduce the energy transfer from the donor fluorophore to the acceptor fluorophore, see Fig.~\ref{fig:FRET}b. 
Changes in the FRET efficiency can be quantified by various methods, including measuring the fluorescence lifetime of the donor fluorophore, i.e. a high FRET efficiency leads to a shortened donor fluorescence lifetime, while a low FRET efficiency leads to an unquenched donor fluorescence lifetime \cite{krainer2015, hellenkamp2018}. Therefore, we quantified the fluorescence lifetimes of the donor in control and tension-reduced conditions (co-incubation with Y27632, 5~$\mu$M) for the cortex-located filamin A molecules with a stretch-sensitive FRET pair (force sensors), see Fig.~\ref{fig:FRET}c. In addition, fluorescence lifetimes were analyzed for the isolated donor molecule and for cortex-located filamin A molecules with a FRET pair located outside the force-bearing elements (zero-force control), see Fig.~\ref{fig:FRET}a and Materials and Methods. The fluorescence lifetime showed multi-exponential behavior, therefore we first fitted the obtained lifetimes with a two-exponential function to determine a short and a long fluorescence lifetimes, $\tau_1$ and $\tau_2$, and the fraction $f_1$ of the first component, see Fig.~\ref{fig:FRET}d1-d3 and Fig.~\ref{fig:S_FLIM-FRET_fit}a. Already, the donor-only data shows two lifetime components, in agreement with earlier reports\cite{fredj2012}. Interestingly, we further saw that the short lifetime component $\tau_1$ remains nearly constant, while the longer lifetime component $\tau_2$ is affected by different tension states.

To be able to deduce the fraction of stretched FRET pairs in cortex-located filamin A molecules, we decided to analyse FRET fluorescence lifetime decays with a more refined model (see Fig.~\ref{fig:FRET}e), considering 4 photophysical states of the donor fluorophore:  In our model , the donor mCerulean can be in two conformations associated with two distinct radiative relaxation rates $k_{r1}$ and $k_{r2}$. Donors in each photophysical conformation are divided into two sub-populations --- one that can undergo FRET with rate $k_e$ (top row) and one that cannot due to large stretch or due to the lack of a mature  acceptor fluorophore (bottom row). We note that the FRET efficiency is assumed to be always the same if FRET occurs. In order to assess parameters of this model, we pursued a workflow as described in Fig.~\ref{fig:FRET}f, \ref{fig:S_FLIM-FRET_fit}b and Materials and Methods. Our results of model fitting show that the fraction $p_e$ of FRET-enabled FRET pairs decreases through tension sensitivity of FRET in particular in cells with high cortical tension, see Fig.~\ref{fig:FRET}g. This observation corroborates the idea of FRET efficiency reduction in response to tension-induced linker stretching, see Fig.~\ref{fig:FRET}g. We denote the fraction of strongly stretched FRET pairs as $f_{st}$. Correspondingly, the fraction of FRET-enabled FRET pairs $p_e$ should decrease by a factor $(1-f_{st})$ in force-sensing conditions. From measured values of $p_e$, we therefore infer a fraction of stretched  FRET pairs in cortex-located filamin A of $f_{st} = 26 \pm 12\%$  in control conditions and $f_{st} = 15 \pm 12\%$ in tension-reduced conditions.  This estimation is in excellent agreement with our previous fit of FRAP and cell deformation experiment data which predicts corresponding fractions of cross-linking filamin A of $f_{cl}=31\%$ and $f_{cl}=13\%$, respectively, in the cortical area analyzed here.

\subsection{The composition of the mitotic cortex shows characteristic changes in response to transient peaks in active tension}
\begin{figure}[ht]
\centering
\includegraphics[width=0.9\linewidth]{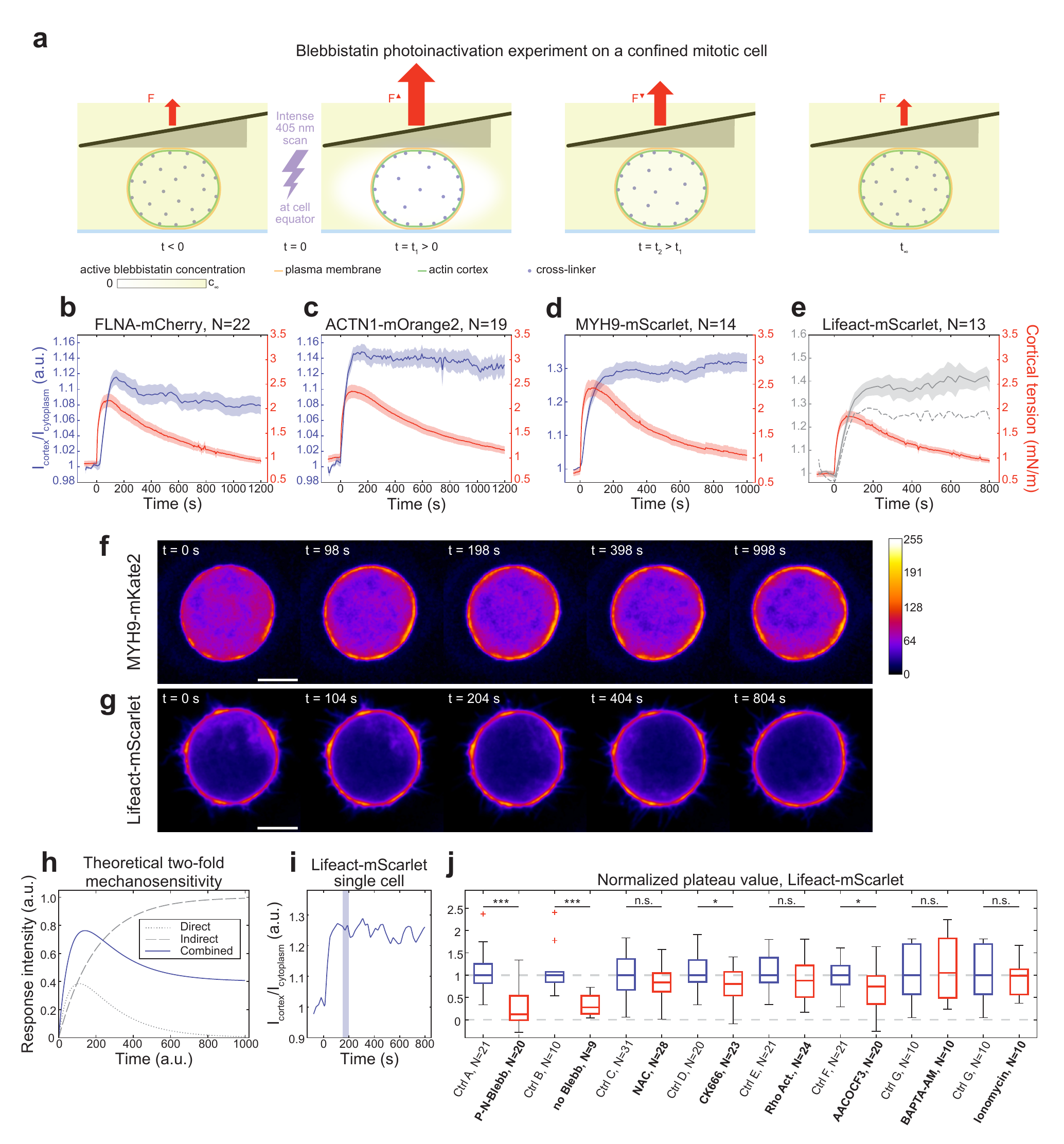}
\caption{\label{fig:BlueLight_1}
Cross-linkers respond to active tension peaks with a twofold mechanosensitive response.
(a) Schematic of experimental procedure of photoinactivation of myosin II inhibitor blebbistatin in confined mitotic cells with blue light irradiation. 
(b-e)  Time evolution of cortical tension and cortex-to-cytoplasm ratio obtained in response to inhibitor-photoinactivation. Cells were expressing diverse cortex-binding fluorescently labeled proteins: (b) filamin A, 
(c) $\alpha$-actinin-1, (d) myosin II, (e) Lifeact (marker of f-actin).
Solid curves (cortex-to-cytoplasm ratio: grey or blue, cortical tension: red) show the mean value and shaded areas indicate the standard error of the mean.  In panel e, the grey dashed line shows the cortex-to-cytoplasm ratio, where the cytoplasmic intensity is measured in the outer shell of the cytoplasm.
(f,g) Time series of confocal images of the equatorial cross-section of cells expressing fluorescently labeled (f) myosin II and (g) Lifeact (marker of f-actin) during experiments with inhibitor-photoinactivation. Time $t=0\,$s corresponds to the time point when photoinactivation was started through light exposure.
Scale bars: 10 µm. The color scale of pixel values is common to both panels. 
The range of pixel intensity values was adjusted and a Gaussian blur filter of radius $0.2~\mu$m was applied to make the time dynamics more visible. No bleaching correction was applied. 
(h) Model of a superposition of two kinds of mechanosensitive responses to an active tension peak. The direct response of the cross-linkers' binding (gray dotted line) combines with the indirect response of f-actin recruitment at the cortex (gray dashed line), and together they account for the observed evolution of the cortex-to-cytoplasm ratio measured in cross-linkers (blue solid line). In the combined curve, the amplitude prefactors of the direct and indirect response were chosen as $1.5$ and $0.4$, respectively, see Materials and Methods.
(i) Exemplary time evolution of cortex-to-cytoplasm ratio of Lifeact-mScarlet without cell confinement (no AFM present). The blue shaded area shows the time interval in which the intensity ratio increase was evaluated (see panel j).
(j) F-actin increase at the cortex after active tension peaks with or without inhibitors or activators of cytoskeletal signaling present (as indicated). Control conditions include $2.5~\mu$M blebbistatin. Values were normalized against the median value of controls in the respective replicate. P-values are calculated using the Wilcoxon rank sum test.
}
\end{figure}
After studying the binding behavior of actin cross-linkers in steady-state conditions and in response to a deformation-induced peak in passive tension, we also wanted to investigate the response of cortex composition to a peak in active tension.
For this purpose, we induced a transient increase in myosin motor activity in a confined mitotic cell co-incubated with blebbistatin. To this end, the cell was exposed to a short pulse of intense blue light, see Fig.~\ref{fig:BlueLight_1}a and Materials and Methods. This treatment locally inactivates blebbistatin and thereby re-activates myosin II \cite{kolega2004}. Afterwards, active blebbistatin progressively diffuses from the outer medium into the cell, leading to a return to the initial tension-reduced state after about 20 minutes. It is noteworthy that the shape of the cell remains the same during the whole process, see Fig.~\ref{fig:BlueLight_1}a,f,g.

We measured the time evolution of the cortex-to-cytoplasm fluorescence intensity ratio following this blebbistatin photoinactivation event in cells expressing fluorescently-labeled filamin A, $\alpha$-actinin-1 and myosin II, see Fig.~\ref{fig:BlueLight_1}b-d, f and Fig.~\ref{fig:S_FRAP_img}d. Strikingly, the cortical localization of each of these cross-linkers increases by more than $10\%$, and even by $30\%$ in myosin II, following the peak in active cortical tension. However, different from what would be expected for the response of a catch-binding cross-linker, the cortex-to-cytoplasm ratio of observed cortical molecules does not relax back to its initial value upon tension relaxation. For filamin A, there is a partial relaxation of the cortex-to-cytoplasm ratio. However, for $\alpha$-actinin-1 and myosin II there is no obvious decrease within a time span of $800\,$s, see Fig.~\ref{fig:BlueLight_1}c-d, f and Fig.~\ref{fig:S_BlueLight_plots}c.

In order to explain this result, we performed the same inactivation experiment on cells expressing fluorescently labeled Lifeact, see Fig.~\ref{fig:BlueLight_1}e and g. We found that the cortex-to-cytoplasm ratio of f-actin increases in a step-like manner by more than $35\%$ in response to an active tension peak.
This result points toward the existence of a second type of mechanosensitivity in the actin cortex present in addition to mechanosensitive binding of cross-linkers to f-actin. This second mechanosensitivity causes a long-term increase of the cortex-to-cytoplasm ratio of f-actin after a peak in active cortical tension in the mitotic cortex.

We noted that the f-actin cortex-to-cytoplasm ratio seems to slowly increase further following the initial fast increase of the first $100$~s, see Fig.~\ref{fig:BlueLight_1}e. In order to investigate that trend, 
we also calculated an alternative cortex-to-cytoplasm ratio where the cytoplasm intensity is measured in the outer rim of the cytoplasm next to the cortex thus excluding the area with chromosomes and microtubules, see grey dashed lines in Fig.~\ref{fig:BlueLight_1}e and \ref{fig:BlueLight_2}b, or grey solid lines in Fig.~\ref{fig:S_BlueLight_plots}a and b. With this alternative definition, the cortex-to-cytoplasm ratio of f-actin only shows a plateau response to the active tension peak without subsequent slow increase. Thus, the slow long-term increase shown in Fig.~\ref{fig:BlueLight_1}e is associated to a decrease of f-actin near the chromosomes. As this cytoplasmic change of f-actin is irrelevant to our study of cortical composition, it is not further addressed in this study. 

The observed long term increase in cortical f-actin in response to active tension peaks can account for the concomitant long-term increase of cross-linkers at the cortex and combines with transient changes of cross-linker binding due to mechanosensitivity in their binding kinetics. 
We therefore put forward that two kinds of mechanosensitivity are at work in combination: i) direct mechanosentivity due to catch-binding and ii) indirect mechanosensitivity due to increased actin polymerization. Fig.~\ref{fig:BlueLight_1}h illustrates this hypothesis by showing a linear combination (blue line) of a step-wise increase due to indirect mechanosensitivity (grey dashed line) and a transient peak due to direct mechanosensitivity (grey dotted line). We note that it resembles the measured response in filamin A shown in Fig.~\ref{fig:BlueLight_1}b. Increasing the weight of the indirect mechanosensitivity in this combination changes the resulting curve such that it becomes similar to the response of $\alpha$-actinin-1 shown in Fig.~\ref{fig:BlueLight_1}c.

To identify molecular mechanisms that are at the heart of the indirect mechanosensitivity, we performed a small screen using pharmacological inhibitors/activators of cortical regulators. For that purpose, we evaluated the relative increase of Lifeact cortex-to-cytoplasm ratio 150-200~s after photoinactivation, see Fig.~\ref{fig:BlueLight_1}i and j. We find that the inactivation-induced increase of f-actin is mostly removed if an increase of cortical tension is prevented by additional co-incubation with the photostable myosin II inhibitor para-nitroblebbistatin, or in the absence of blebbistatin (first 2 conditions in Fig.~\ref{fig:BlueLight_1}j). In order to test whether the release of Reactive Oxygen Species (ROS) plays a role for the observed effect, we incubated cells with the ROS scavenger N-acetyl-L-cysteine (NAC) during the experiment. We find that NAC does not inhibit f-actin increase, suggesting that ROS release is not responsible for the observed response. Testing the influence of actin nucleators Arp2/3 and RhoA via treatment with the Arp2/3 inhibitor CK666 or the Rho Activator II, we find that Arp2/3 activity contributes to the observed f-actin increase while Rho activation appears to play no major role. Furthermore, previous work reported cortical f-actin increase upon cell and, in particular, nuclear squeezing in interphase cells, triggered via phospholipase A2 signaling and calcium release from internal membrane stores \cite{lomakin2020, venturini2020}. We therefore also tested the influence of the phospholipase A2 inhibitor AACOCF3, of the cytoplasmic calcium chelator BAPTA-AM, and of ionomycin which permeabilizes cell membranes for calcium. Surprisingly, we find that neither BAPTA-AM nor ionomycin affect the increase of f-actin, whereas the phospholipase A2 inhibitor partially inhibits it. We conclude that calcium signaling is not a major factor of the indirect mechanosensitivity. We speculate that the impact of phospholipase A2 might be through previously reported interaction between myosin motors and internal cell membranes such as the Golgi membranes \cite{de_figueiredo1998,liu2016}. 
Previous studies showed that Golgi membranes are enriched in Arp2/3 activators \cite{egea2015}. 
Thereby, mechanical stimulation of internal membranes via activated myosin II motors might play a role in enhanced actin polymerization.
The exploration of the molecular mechanisms of this stimulation via myosin II is an interesting topic in itself, which we leave for future investigation.

\subsection{Passive tension peaks prior to active tension peaks partially inhibit indirect mechanosensitivity of f-actin and cross-linkers}
\begin{figure}[ht]
\centering
\includegraphics[width=1\linewidth]{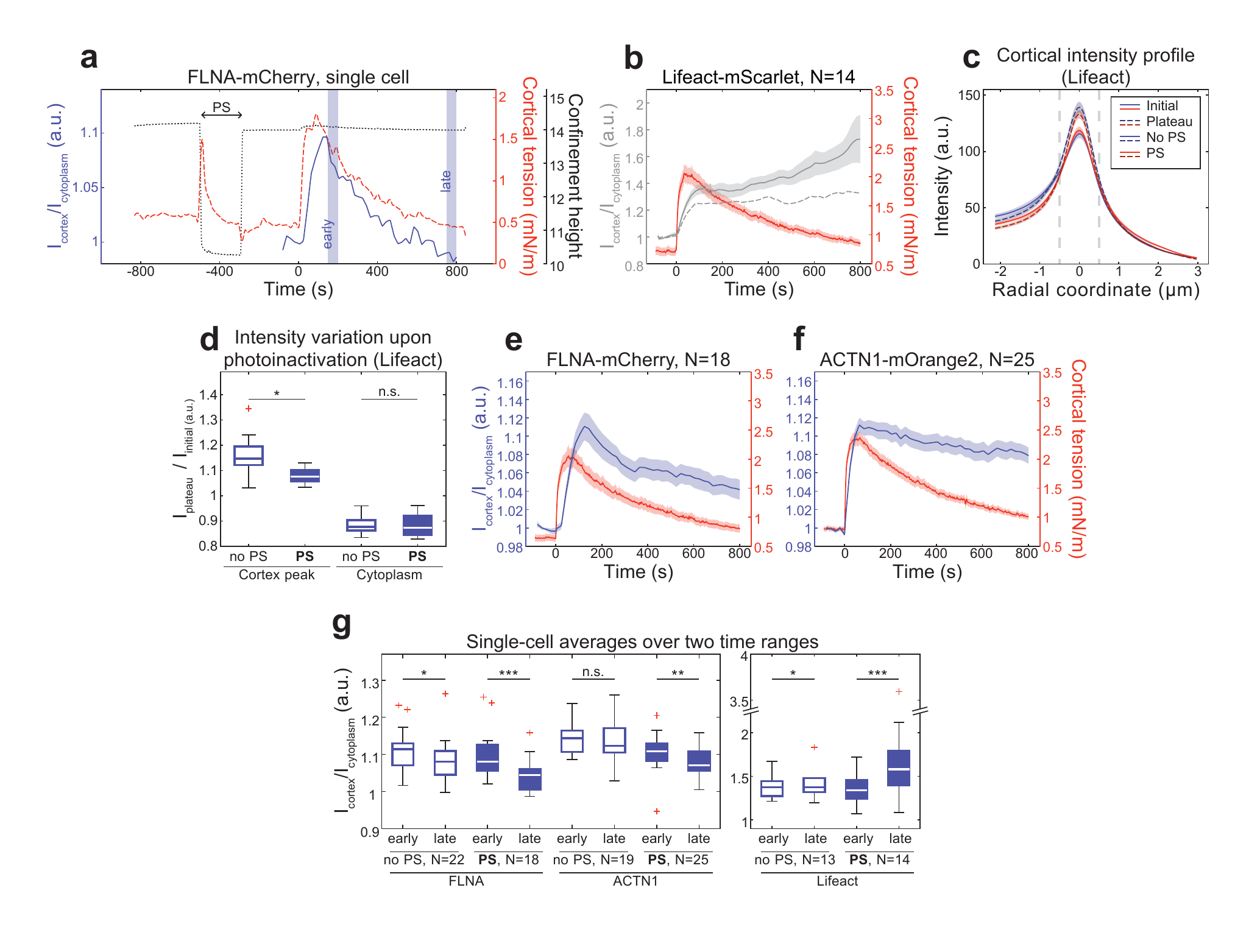}
\caption{\label{fig:BlueLight_2}
Imposing a passive tension peak prior to the active tension peak partially inhibits the indirect mechanosensitive response of f-actin and cross-linkers at the cortex. 
(a) Exemplary time evolution of cortical tension, cantilever height and cortex-to-cytoplasm ratio for one cell during pre-squeezing (PS) and subsequent inhibitor-photoinactivation. The blue shaded areas show the early and late time intervals in which the intensity ratios are evaluated in panel g.
(b) Time evolution of cortical tension and cortex-to-cytoplasm ratio obtained in response to pre-squeezing and inhibitor-photoinactivation in cells expressing fluorescently labeled Lifeact. Shaded areas indicate the standard error of the mean. The grey dashed line shows the intensity ratio where the cytoplasmic intensity is measured in the outer shell of the cytoplasm.
(c) Cell-averaged radial intensity profiles around the cortex of cells expressing fluorescently labeled Lifeact. (Negative arguments correspond to the cell interior.) Profiles are shown before inhibitor photoinactivation (solid lines) and 150-200~s after photoinactivation (dashed lines, plateau in panel b). Cells were measured either with pre-squeezing (red lines, N=14) or without (blue lines, N=13). Shaded areas represent the standard error of the mean. Grey dashed lines indicate the region of the cortex intensity peak used in panel d.
(d) F-actin increase at the cortex upon photoinactivation is diminished through pre-squeezing. Shown are boxplots of the ratio between the increased intensity 150-200~s after inhibitor-photoinactivation and the initial intensity, in the cortex intensity peak ($1~\mu$m thickness, shown in panel c and in the cytoplasm.  Cells were measured either without pre-squeezing (empty boxes, N=13) or with pre-squeezing (solid boxes, N=14).
(e,f) Time evolution of cortical tension and cortex-to-cytoplasm ratio obtained in response to pre-squeezing and inhibitor-photoinactivation. Shaded areas indicate the standard error of the mean.
(g) Boxplots of normalized intensity ratios at the early and late time points (see panel a) for fluorescently labeled  filamin A, $\alpha$-actinin-1 and Lifeact with and without pre-squeezing (PS).
P-values are calculated using the Wilcoxon rank sum test.
}
\end{figure}

Verifying the presence of a direct mechanosensitive response after an active tension peak is challenging due to its superposition with the indirect mechanosensitive response, see Fig.~\ref{fig:BlueLight_1}h.
Therefore, we sought for a way to diminish the indirect cortical mechanosensitivity in this experiment. Since none of the pharmacological treatments we applied could abandon indirect mechanosensitivity while keeping direct mechanosensitivity, see Fig.~\ref{fig:BlueLight_1}j, we tried to inhibit the indirect mechanosensitivity by means of a preceding mechanical stimulation. 
For this purpose, we designed a new experiment where the cell deformation procedure shown in Fig.~\ref{fig:Squish}a is performed before the blebbistatin photoinactivation procedure, see Materials and Methods. 
Prior to the photoinactivation event itself, the cantilever is  lowered to squeeze the confined cell and then lifted back to its normal confinement position -- a sequence that we term ``pre-squeezing'', see Materials and Methods. Sufficiently long time periods are left to allow for relaxation of both the tension and the direct mechanosensitive response of the cross-linkers between the two cantilever movements and between the pre-squeezing (noted PS) and the photoinactivation event at $t=0~$s.
In Fig.~\ref{fig:BlueLight_2}a, corresponding cell confinement height (black dotted line), cortical tension (red dashed line) and cortex-to-cytoplasm ratio of filamin A (blue solid curve) are shown for this experiment over time for an exemplary cell. The two blue shaded areas indicate the time ranges we used to compare the response of the cortex molecules to the active tension peak at early and late times.

While the cortex-to-cytoplasm ratio of f-actin increases by $\approx 35\%$ in response to active tension peaks with or without pre-squeezing, see grey curves in Fig.~\ref{fig:BlueLight_1}e and \ref{fig:BlueLight_2}b, we also separately examined the change of cortical and cytoplasmic intensities of the fluorescent f-actin marker upon active tension peaks; we found that radial intensity profiles of Lifeact change upon blebbistatin photoinactivation on average (before: solid lines, early plateau: dashed lines in Fig.~\ref{fig:BlueLight_2}c) such that f-actin at the cortex increases and f-actin in the cytoplasm decreases. This redistribution of f-actin is modified through pre-squeezing (no pre-squeezing: blue curves, pre-squeezing: red curves in Fig.~\ref{fig:BlueLight_2}c); in particular, the f-actin increase at the cortex is diminished, likely because f-actin increase had been partially anticipated through pre-squeezing. 
Calculating the integral of fluorescence intensity over a $1~\mathrm{\mu m}$ interval centered around the peak cortical intensity (see grey dashed lines in Fig.~\ref{fig:BlueLight_2}c), we quantified f-actin changes as the ratio of integral values after and before photoinactivation.
We found that f-actin in this cortex region increases significantly less if pre-squeezing is performed, see Fig.~\ref{fig:BlueLight_2}d. By comparison, the change of fluorescence intensities in the cytoplasmic bulk is unaltered by pre-squeezing, see Fig.~\ref{fig:BlueLight_2}d. We conclude that adding mechanical stimulation through pre-squeezing to our protocol partially inhibits the actin polymerization boost at the cortex in response to active tension peaks and, thus, diminishes the effect of indirect mechanosensitivity. In accordance with this, we observed that the direct mechanosensitive response of $\alpha$-actinin-1 and  filamin A is exposed more with preceding mechanical stimulation, see Fig.~\ref{fig:BlueLight_2}e-g.

\section{Discussion}
In this work, we studied the mechanosensitivity of the actin cortex using mitotic HeLa cells as cellular model system. In particular, we were interested in structural and compositional changes of the cortex in response to peaks of active or passive mechanical tension in the cortex. Beyond the major cortical building blocks actin and myosin II, we focused  specifically on changes in cortical association and binding dynamics of essential actin cross-linkers since actin cross-linkers are important regulators of cortical mechanics. %
For this purpose, we used a previously established assay where round mitotic cells are brought into parallel plate confinement via the cantilever of an atomic force microscope. This confinement assay allows for controlled cell deformation and a continuous readout of cortical tension during the experiment \cite{hosseini2020}. At the same time, 
cortex-associated proteins were imaged with a confocal microscope to monitor the time evolution of cortical association.

Firstly, we examined the effect of cortical tension on molecular turnover and the cortical localization of cross-linkers.
We show that $\alpha$-actinin-1 and filamins A and B exhibit hallmarks of catch-binding behavior in this assay: recovery timescales after photobleaching and fractions of cortex-associated molecules increase with increasing tension, see Fig.~\ref{fig:FRAP}f,g. Interestingly, for myosin II, we find no signs of tension-associated changes, see Fig.~\ref{fig:FRAP}f,g. 

Investigating further, we probed the effect of peaks in passive deformation-induced tension on the cortical localization of our cross-linkers of interest. For this purpose, we used an experiment where a quick deformation is imposed on the cell to induce the tension peak, see Fig.~\ref{fig:Squish}a-g. We observed that the induced peak in passive tension was linked to a subsequent peak in the cortical association of filamins A and B shortly after.
This observation provides additional support that these cross-linkers form catch-bonds with f-actin. In this scenario, a reduced unbinding rate ($\hat =$ longer bond lifetimes) in the phase of higher tension leads to an enhanced growth of the cortical cross-linker population and a subsequent cortical concentration peak. 

Next, we fitted a simple kinetic binding model for mechanosensitive cross-linkers to the data obtained from our FRAP experiments and cell-squeezing experiments as well as to published data on $\alpha$-actinin-4 {} \cite{hosseini2020}, see Table~\ref{tab:Tab1}. The model includes a tension-dependent unbinding rate for cross-linkers simultaneously bound to two actin filaments, see Fig.~\ref{fig:FRAP}a. This is motivated by the fact that only such cross-linkers can be mechanically loaded. Tension sensitivity of this unbinding rate is scaled by the model parameter $\alpha$, where a positive value corresponds to catch-binding, see Eq.~\eqref{eq:TaylorExpRate}.  We find that $\alpha$ is significantly larger than zero for $\alpha$-actinins 1 and 4, as well as filamins A and B. Correspondingly, the results of our FRAP experiments and cell deformation experiments taken together strongly support the hypothesis that $\alpha$-actinins 1 (and 4), and filamins A and B form catch bonds with f-actin. Interestingly, the mechanosensitivity parameter $\alpha$ is about one order of magnitude higher in filamins than in $\alpha$-actinins, although these cross-linkers all have similar actin binding sites constituted by a pair of N-terminally located calponin homology domains \cite{harris2019,gimona2002,weins2007}.
In particular, we see only a very shallow mechanosensitivity of $\alpha$-actinin 1 which exhibits the fastest binding kinetics, see Fig.~\ref{fig:FRAP}f and \ref{fig:ViolinPlots}a-d. 
We speculate that this observation might be due to the fact that cross-linkers need to stay in a cross-linking state long enough (compared to f-actin turnover) to become mechanically loaded, making sufficiently slow binding kinetics a prerequisite for impactful catch-binding in cross-linkers.

In order to provide a benchmarking experiment with an independent method, we performed FLIM-FRET measurements on filamin A constructs containing a FRET molecular force probe. We measured the steady-state fraction of cortex-associated filamins that get stretched by cortex-internal forces arising from the cortical tension. We find that this fraction of stretched FRET force probes increases with increasing cortical tension in a way that is consistent with our quantitative predictions of kinetic binding obtained from FRAP and cell-squeezing experiments. 

Finally, we examined the response of the cortical localization of $\alpha$-actinin-1, filamin A and myosin II, and of the amount of cortical f-actin to peaks in active tension induced by short term activation of myosin II molecular motors. Surprisingly, we found that the amount of cortical f-actin increases in response to active tension peaks in the mitotic cortex and remains elevated over a long time, see Fig.~\ref{fig:BlueLight_1}e and g.
We identify this observation as a second, indirect kind of cortical mechanosensitivity, which is present in addition to the direct mechanosensitivity that we had observed so far in the binding of actin cross-linkers. This combination of a direct and an indirect mechanosensitive responses is reflected in our measurements with the three cross-linkers, see Fig.~\ref{fig:BlueLight_1}b-g. 
Moreover, we found that applying a mechanical stimulation to the cell right before peaks in active tension mitigates the observed effect of indirect mechanosensitivity and better reveals the direct mechanosensitive response of the cross-linkers $\alpha$-actinin-1 and filamin A by exposing a short term peak in their cortex-to-cytoplasm ratio right after the tension peak, see Fig~\ref{fig:BlueLight_2}e-g. 

In contrast to our results on $\alpha$-actinins and filamins, we observed no hallmarks of (direct) mechanosensitivity in the f-actin binding dynamics of myosin II in our experiments. This came as a surprise given that extensive research has shown that tensile forces lead to an increase in the lifetime of bonds of myosin II heads with actin \cite{kovacs2007, norstrom2010}. This apparent discrepancy between our results and the literature likely originates from the molecular details of myosin II binding to f-actin;  myosin II molecules bundle into a mini-filament which then binds to f-actin with several myosin heads at once \cite{niederman1975, verkhovsky1993}. 
Therefore, the cortical lifetime of a mini-filament is sufficiently longer than the binding time of an individual myosin head \cite{grewe_mechanosensitive_2020} and may exceed several hundreds of seconds \cite{bennett_smooth_2022}. In this way, short-lived tension peaks, as applied in our assay, may be too short to significantly affect myosin mini-filament concentration at the cortex. Alternatively, the applied tension increase may not have been high enough to change myosin binding in a meaningful manner. 
We suggest that the observed relatively fast fluorescence turnover of myosin II at the cortex reflects intramolecular turnover of bound mini-filaments rather than binding and unbinding events of mini-filaments to and from the cortex. In fact, intramolecular mini-filament turnover  has been recently reported to be on a similar time scale as our FRAP turnover time, i.e. $\approx 50\,$s, see Fig.~\ref{fig:FRAP}f and {} \cite{bennett_smooth_2022}. 
Furthermore, we note that (para-nitro)blebbistatin, the reagent that we use to manipulate cortical tension, interferes with myosin II binding kinetics and may thereby affect myosin II cortical association in ways that are not captured by our model \cite{kovacs2004}. 

Previous research has pointed at a mechanosensitivity of cross-linker proteins of the cytoskeleton. Luo \textit{et al.} and Schiffhauer \textit{et al.} \cite{luo2013, schiffhauer2016} showed that cross-linkers $\alpha$-actinin, filamin as well as myosin II accumulate in regions of the cortex subject to external deformation via micro-pipette aspiration. In these studies, monitoring of cortical cross-linker dynamics was not combined with measurements of cortical tension dynamics. Therefore, it remains unclear if and how observed changes in cortical composition are related to changes in cortical tension \cite{belly_actin-driven_2022}. 
In particular, mechanical deformation steps of the cortex as applied by micropipette aspiration only give rise to short term changes in cortical tension due to the viscoelastic nature of the cortex material \cite{bonfanti2020, fischer-friedrich2016, hosseini2021}. Therefore, the long term change described in {} \cite{luo2013, schiffhauer2016} in the binding dynamics of actin cross-linkers cannot be accounted for by induced changes in cortical tension but likely originates from a different kind of cellular mechanosensitivity such as curvature-dependent activity of actin nucleators which has previously been reported e.g. for Arp2/3 {} \cite{pipathsouk_wave_2021, takano_efcf-bar_2008}. 

In light of all our experimental results, we conclude that the cortex is a smart self-tuning material that reinforces itself through structural changes in response to peaks in mechanical tension. We find that this response incorporates two kinds of mechanosensitivity --- i) a direct mechanosensitivity through force-sensitive binding of cortical constituents and ii) an indirect mechanosensitivity 
which leads to changes in the polymerization dynamics of actin. 
According to our analysis, direct cortical mechanosensitivity is, at least in part, mediated by catch-binding  of several actin cross-linkers, in particular $\alpha$-actinin-1 and 4, filamin A, as well as filamin B featuring the cortex as more cross-linked at higher tensions, see Fig.~\ref{fig:ViolinPlots}e,f. 
Cross-linking density has been shown to be a parameter that sensitively tunes the stiffness of semi-flexible polymer networks such as the actin  cytoskeleton and an increase in cross-linking likely contributes to frequently reported cytoskeletal stress-stiffening  \cite{gardel2004,gardel2006b,lieleg2009, fischer-friedrich2016,stam04,wang02,fern08}.
The here reported indirect mechanosensitivity response of the cortex is particularly prominent in response to active tension peaks.
It gives rise to increased actin polymerization and increased myosin II at the cortex which reinforces the cortex mechanically in addition to enhanced cross-linking. 

Finally, we note that cortical cross-linkers and f-actin are known mediators of extra- and intracellular signaling \cite{nakamura2011, kanchanawong_organization_2023, chalut2016} such that changes of their cortical concentration upon tension peaks not only give rise to mechanical changes due to their structural function but are also expected to elicit essential changes in cellular signaling. 

\section{Materials and Methods}
\subsection{Cell culture}
We cultured the HeLa Kyoto cells in Dubelcco's modified Eagle's medium (DMEM, \#31966-021, Life Technologies) supplemented with $10\%$ v/v fetal bovine serum (FBS, \#10270106, Life Technologies) and $1\%$ v/v penicillin/streptomycin (\#15140122, Life Technologies).
Cell culture flasks were kept at 37°C and under $5\%$ $\mathrm{CO_2}$. Cells were split every 2 to 3 days when they reached 60 to $80\%$ confluency.
Transfected cells were cultivated in selection media containing 400 $\mu$g/ml G418 sulfate (geneticin, \#10131035, Life Technologies).

Pharmacological agents were added from stock solutions at least 15~min prior to measurements to indicated concentrations.
The following DMSO stock solutions were used:  (-)-blebbistatin $10~\mathrm{mM}$ (\#B0560-1MG, Sigma), para-nitroblebbistatin $20~\mathrm{mM}$ (P-N-Blebb, \#DR-N-111, Optopharm), Rock inhibitor Y-27632 $10~\mathrm{m M}$ (Y27, \#1005583, Cayman Chemical), Arp2/3 inhibitor CK-666 $20~\mathrm{mM}$ (\#29038, Cayman Chemical), BAPTA-AM $30~\mathrm{mM}$ (\#15551 Cayman Chemical), ionomycin $5~\mathrm{mM}$ (\#sc-3592, Santa Cruz Biotechnology), S-trityl-L-cysteine $10~\mathrm{mM}$ (STC, \#164739, Sigma). Moreover, we prepared a stock solution of arachidonyl trifluoromethyl ketone (AACOCF3, \#ab120350, Abcam) at $5~\mathrm{mg/mL}$ in ethanol, a stock solution of N-acetyl-L-cysteine (NAC, \#A7250, Sigma-Aldrich) at 100 mM pH-adjusted and in PBS, and a stock solution of Rho Activator II (Rho Act., \#CN03, Cytoskeleton Inc.) at $0.1~\mathrm{\mu g/\mu L}$ in water.

\subsection{Transfection}
Stable HeLa cell lines expressing FLNA-mCherry (Addgene \#55047), Lifeact-mScarlet (Addgene \#85054) or ACTN1-mOrange2 (Addgene \#57944 ) were generated through chemical transfections using Turbofectin 8.0 (\#TF81001, OriGene) or Lipofectamine 2000 (\#11668019, Invitrogen) according to the manufacturer's instructions. 
The vector pLifeAct\_mScarlet\_N1 was a gift from Dorus Gadella (Addgene plasmid \#85054; \url{http://n2t.net/addgene:85054}; RRID:Addgene\_85054). 
The vectors mOrange2-Alpha-Actinin-19 (Addgene plasmid \#57944; \url{http://n2t.net/addgene:57944}; RRID:Addgene\_57944), and mCherry-FilaminA-N-9 (Addgene plasmid \#55047; \url{http://n2t.net/addgene:55047}; RRID:Addgene\_55047) were gifts from Michael Davidson. 

HeLa cell lines expressing FLNA-GFP, FLNB-GFP and MYH9-mKATE2 were generated by transfection with bacterial artificial chromosomes using Effectene (\#301425, Qiagen) as described in Poser {\it et al.}\cite{poser2008}.

Transient transfections for expression of FLNA-M-stFRET (force-sensing FLNA construct), FLNA-C-stFRET (zero-force construct) and mCerulean (FRET donor) were performed with Turbofectin 8.0. Transfection complexes were added to polystyrene multiwell plates along with freshly detached cells 40 to 45 hours prior to the measurements. The transfection medium was removed after 24 hours. %
The vectors Filamin-M-stFRET (Addgene plasmid \#61105, \url{http://n2t.net/addgene:61105}, RRID:Addgene\_61105) and Filamin-C-stFRET (Addgene plasmid \#61107, \url{http://n2t.net/addgene:61107}, RRID:Addgene\_61107) were  gifts from Fred Sachs. The vector mCerulean-N1 was a gift from Michael Davidson \& Dave Piston (Addgene plasmid \#54758, \url{http://n2t.net/addgene:54758}, RRID:Addgene\_54758).

\subsection{Experiments using both AFM and confocal microscopy}
\label{sec:M&M_AFM}
One day before the experiments, we seeded the cells into silicon cultivation chambers ($0.56~\mathrm{cm^2}$, from ibidi 12 well chamber) placed in 35-mm glass-bottom dishes (FD35-100, Fluorodish). Two to eight hours before the measurements, we removed the silicon inserts and replaced the growth medium by $\mathrm{CO}_2$-independent DMEM (\#12800-017, Invitrogen) with 4 mM NaHCO3 buffered with 20 mM HEPES/NaOH pH 7.2 and supplemented with 10\% FBS (imaging medium). %
We simultaneously added STC at a concentration of $2~\mathrm{\mu M}$ to induce mitotic arrest. Where indicated, we also added a specific drug to the cells at least $15$ minutes before the measurement start.

We performed the experiments described in this paragraph with an LSM 700 or an LSM 510 confocal microscopes (Zeiss) of the CMCB light microscopy facility at TU Dresden, in combination with either a Nanowizard 1 or a Nanowizard 4 AFM (JPK Instruments). 
In either case, we used a Plan Apochromat 20x 0.8 objective (Zeiss), and a $488$ or $555~\mathrm{nm}$ excitation laser. During the measurement, cells were kept at 37°C using a petri dish heater (JPK instruments).

We prepared tipless AFM cantilevers (HQ:CSC37/tipless/No Al, with a nominal spring constant $0.3$ to $0.8~\mathrm{N/m}$, Mikromasch) for the parallel confinement assay by adding a wedge made of UV-curing adhesive (Norland Optical Adhesive 63, Norland Products) in order to correct their 10°-tilt \cite{stew13}. 
On every measurement day, the spring constant of the cantilever was calibrated using the thermal noise analysis (built-in software, JPK). 

For each single-cell measurement, we first approached the cantilever to the dish bottom in order to measure its relative z position. We then retracted it by about $14.5~\mathrm{\mu m}$, and gently brought it over a mitotic cell to achieve its parallel plate confinement. We allowed the force applied by the cell to the cantilever to relax before the next steps. We set up the confocal microscope acquisition for frames of $256\times 256$ or $512\times 512$ pixels covering a field of $30\times 30$ to $45\times 45~\mathrm{\mu m}$ that contained the mitotic cell, focusing at its equatorial cross-section.

We continuously recorded the force on the cantilever and its height relative to the dish bottom using software from the AFM's manufacturer (JPK Instruments). Then, we used these data together with the confocal images to determine the cell's surface area, its volume, and its cortical tension as described previously \cite{hosseini2020}.

\subsubsection{Fluorescence recovery after photobleaching (FRAP)}
FRAP experiments on the cortex of cells expressing fluorescent cross-linkers were done as in Hosseini {\it et al.} \cite{hosseini2020}. Briefly, we selected a ROI covering about $10\%$ of the circumference of the cell and quickly photobleached it using repeated scans of the excitation laser at full power. We imaged the whole frame five times before and 150 to 250 times after the bleaching, with a time interval of $0.5$ to $4$~s. We later analyzed the time series 
i) extracting the intensity in the bleached region of the cortex over time, 
ii) normalizing this intensity by its initial value before bleaching, and 
iii) correcting fluorescence intensities for bleaching upon continuous imaging. 
When indicated, we treated the cells with $2.5~\mathrm{\mu M}$ blebbistatin or $10~\mathrm{\mu M}$ para-nitroblebbistatin prior to the  measurement. 
The matlab script we used for analysis of FRAP data is publicly available under \url{https://gitlab.com/polffgroup/actincortexanalysis}.

\subsubsection{Cell-squeezing experiments}
Cell-squeezing  experiments were done as in Hosseini {\it et al.} \cite{hosseini2020}. Briefly, we treated the cells with $5~\mathrm{\mu M}$ of the Rock inhibitor Y27632 in order to keep the cells in a reference state of low cortical tension. For experiments with cells expressing MYH9-mKate2, the cortex-relaxing drug  blebbistatin ($5~\mathrm{\mu M}$) was used instead of the Rock inhibitor Y27632. This was done since Rock inhibition leads to a significant reduction of cortical myosin and thereby to a strong reduction of the signal-to-noise ratio in the experiment. 
Drugs were added at least 15 min prior to the measurement.

During the measurement, we placed a selected mitotic cell under AFM confinement and waited for the AFM force to stabilize as described above. Then, we lowered the wedged cantilever by $2$ to $5~\mathrm{\mu m}$ at a speed of $0.5~\mathrm{\mu m/s}$ depending on the cell line, where the maximum height change was applied that did not yet trigger blebbing (FLNA: $5~\mu$m, FLNB: $5~\mu$m, Lifeact: $4~\mu$m, ACTN1: $3~\mu$m, and MYH9: $2-3~\mu$m). For experiments with cells expressing MYH9-mKate2, the cantilever was lowered at speeds of 0.2 to $0.3~\mathrm{\mu m/s}$.

We started to image the equatorial cross-section of the squeezed cell immediately after this height ramp. 
We then analyzed the obtained time series of images to extract the fluorescence intensities at the cortex and in the cytoplasm over time, correcting for the background intensity as measured in the vicinity of the cell. Finally, we plotted the corrected cortical intensity divided by the corrected cytosolic intensity over time together with the cortical tension. The script we used for analysis of cell-squeezing data is publicly available under \url{https://gitlab.com/polffgroup/actincortexanalysis}.

\subsubsection{Blue light photoinactivation of blebbistatin}
Experiments with quick inactivation of blebbistatin by blue laser light were done similar to the work presented in Hosseini \textit{et al.} \cite{hosseini2020}. Briefly, we added $2.5~\mathrm{\mu M}$ blebbistatin to the cell's medium at least $15$ minutes before the experiment. We placed a single mitotic cell under confinement and selected a square ROI, located at the cell's equator and slightly bigger than its cross-section, for blue light illumination. When the AFM force readout had stabilized, we imaged the cell's equatorial cross-section for $80~\mathrm{s}$ with $555~\mathrm{nm}$ excitation. Subsequently, we triggered a repeated scan of the above-mentioned ROI  with a $405~\mathrm{nm}$ laser at a high power using the confocal microscope. This intense blue illumination induced the inactivation of the blebbistatin present in the cell and thus a temporary rise of the myosin contractility. In all plotted experimental results, time $t=0$ corresponds to the beginning of the blue light scan. 
We then measured the cell's cortical tension and imaged the orange- or red-fluorescent cross-linkers at the cell's equatorial cross-section for up to 25 minutes. We set the power of the $405~\mathrm{nm}$ laser and the number of photoinactivation scan repeats so as to induce at least a doubling of the cortical tension with less than $20~\mathrm{s}$ of blue illumination.

We analyzed the time series of images in the same way as for the cell-squeezing experiments, and generated time series of the cortex-to-cytoplasm ratio of the molecule of interest. We also generated time series of an alternative intensity ratio where the cytoplasm intensity used is the integrated intensity over a $0.6~\mu$m-wide circular band in the outer part of the cytoplasm. Using this alternative cortex-to-cytoplasm ratio essentially removes the contribution of the central region of the mitotic cell, where the chromosomes are, from the analysis.

Additionally, we did the same experiment without AFM confinement and imaging 2 to 4 cells at a time. We used this higher-throughput protocol to assess the effect of the following activators or inhibitors of cortical signaling: para-nitroblebbistatin ($20~\mathrm{\mu M}$), CK666 ($5~\mathrm{\mu M}$), Rho Activator II ($2~\mathrm{\mu gm/L}$), arachidonyl trifluoromethyl ketone ($20~\mathrm{\mu M}$), BAPTA-AM ($15~\mathrm{\mu M}$), ionomycin ($2~\mathrm{\mu M}$), and N-acetyl-L-cysteine ($6~\mathrm{mM}$). For each of these conditions, we added the indicated drug to the medium at the same time as blebbistatin. We measured cells in the control condition -- with blebbistatin only -- on the same day for comparison, see Fig.~\ref{fig:BlueLight_1}j. Furthermore, we also performed this experiment without blebbistatin in the medium, see Fig.~\ref{fig:BlueLight_1}j, condition 2.

Furthermore, we also combined the blue light photoinactivation experiment with a prior mechanical stimulation, which we termed ``pre-squeezing'', see Fig.~\ref{fig:BlueLight_2}a. After placing a single mitotic cell under AFM confinement, we waited until the AFM readout had stabilized and imaged subsequently the equator cross-section of the cell. Then, we lowered the wedged cantilever, following the same height ramp as during cell-squeezing experiments on cells of the same line. We kept the cantilever at its lowered position for $200~\mathrm{s}$ and then lifted it back to its original position, at $0.5~\mathrm{\mu m/s}$. After another relaxation period of ca. $200~\mathrm{s}$, we started the blue light blebbistatin photoinactivation protocol, including the initial imaging before the blue light scan. 

\subsection{Acquisition of FLIM-FRET data}
We performed all fluorescence lifetime imaging microscopy with an LSM 780 confocal laser scanning microscope (Zeiss) with a C-Apochromat 40x/1.20 W Corr objective and a Becker \& Hickl time-correlated single-photon counting (TCSPC) module. A $440~\mathrm{nm}$ laser diode pulsed at $80~\mathrm{MHz}$ was used for excitation. The emission signal was reflected on a $505~\mathrm{nm}$ beam splitter and filtered through a $480/40~\mathrm{nm}$ bandpass filter before its acquisition by an HPM-100-40 hybrid GaAsP detector (Becker \& Hickl).

About 16 hours before the measurements, we transferred the transiently transfected cells from their multiwell plates to glass-bottom dishes with normal culture medium. 
To increase the number of mitotic cells, we supplemented the medium of  transfected cells with $2\,\mathrm{\mu M}$ STC for an overnight pre-incubation. Then, 2 to 8 hours prior to imaging, we changed the medium to the same $\mathrm{CO}_2$-independent medium as used in our AFM experiments, with $2\,\mathrm{\mu M}$ STC. 
We kept the cells at 37°C by without $\mathrm{CO_2}$ during the whole experiment.

We selected and imaged single mitotic cells expressing the construct of interest. We focused the microscope on the equatorial plane of the cell, and set a frame of $256\times 256$ pixels and $51\times 51~\mathrm{\mu m}^2$ to be scanned
continuously
until about 1000 photons per pixel were counted in the cell's cortex.

To measure the instrument response function (IRF), we imaged fluorescein dissolved into a saturated potassium iodide solution, with the same acquisition settings as for the cells. Collisional quenching of fluorescein by the highly concentrated iodide ions shortens its fluorescence lifetime down to a few $10~\mathrm{ps}$ \cite{liu2014}. The obtained decay curve is hence a good approximation of the true IRF. We used this fluorescein decay curve as the measured IRF in our analysis procedure.

\subsection{Analysis of FLIM-FRET data}
Using the FLIMfit software (version 5.1.1, Imperial College London, United Kingdom), we segmented the FLIM images of single cells to produce one combined decay curve per cell.
In cells expressing the FLNA stFRET constructs, we selected pixels in the fluorescent cortical rim, i.e. a 1 to $2~\mathrm{\mu m}$-wide band along the cell edge (ca. 1000 to 2000 pixels). 
The lifetime decays of all pixels were pooled in one averaged decay curve.  
In cells expressing mCerulean (donor) only, we selected the cell in the image using a threshold of 500 photons per pixel, and averaged the decay curves from the selected pixels. 

Subsequently, we fitted the derived decay curves, by reconvolution of the IRF with multi-exponential decay functions using a custom written Matlab script. We restricted the analysis to a time window of $8.8~\mathrm{ns}$, fixed across all conditions and starting $\sim 2~\mathrm{ns}$ after the emission peak. We chose so to improve the fit quality by excluding the region of back-scattered laser intensity.

We first considered the following two-exponential decay function
\begin{equation}
I(t)=\frac{f_1}{\tau_1} e^{-t/\tau_1}+\frac{(1-f_1)}{\tau_2} e^{-t/\tau_2} \quad.\label{eq:FRETTwoExp}
\end{equation}
For each measured cell, we convolved Eq.~\eqref{eq:FRETTwoExp} with the IRF and normalized the theoretical decay within the time window of analysis as it was also done for the experimentally derived decay curves. In order to find the most likely estimator of the fit parameters of Eq.~\eqref{eq:FRETTwoExp}, we used maximum likelihood estimation \cite{maus2001}. The fitting parameters were the timescales $\tau_1$ and $\tau_2$, with the corresponding fraction $f_1$, as well as a background intensity and a time shift for the IRF.

Next, we fitted a model with four donor photophysical states on the data measured for each cell expressing one of the FLNA-stFRET constructs, see Fig.~\ref{fig:FRET}e. 
The model makes the following assumptions; the donor molecule can be in one of two possible conformations with two corresponding radiative relaxation rates $k_{r1}$ and $k_{r2}$.
This assumption was motivated by lifetime measurements of isolated mCerulean, whose distribution is well described by two relaxation timescales, see Fig.~\ref{fig:FRET}d and Fredj \textit{et al.}\cite{fredj2012}.
We assume that stretching of the stFRET sensor due to mechanical load beyond a threshold leads to a substantial increase in the distance between donor and acceptor and therefore to the suppression of FRET. This assumption is based on the fact that the $\alpha$-helical linker polypeptide exhibits a length of $5$~nm, close to the Förster radius (i.e. FRET efficiency of $44\%$), in unstretched state \cite{meng2008}. 
It was previously shown that $\alpha$-helices unfold in an all-or-none manner beyond a threshold force leading to an elongation close to their contour length \cite{schwaiger2002, zegarra2009, meng2008}. This elongation of the FRET linker would lead to a drop in FRET efficiency to almost zero. 
Accordingly, we assume that only stFRET sensors that are mature and unstretched  are able to undergo FRET, with a transfer rate $k_e$. We assume that the probability to be in such a FRET-enabled state is $p_e$ and independent of the conformation adopted by the donor.

To fit lifetime distributions within the four-state model, we used the following multi-exponential decay function
\begin{equation}
I(t)=f_{r1} k_{r1} e^{-k_{r1}t}+f_{e1} (k_{r1}+k_e) e^{-(k_{r1}+k_e)t}+f_{r2} k_{r2} e^{-k_{r2}t}+f_{e2} (k_{r2}+k_e) e^{-(k_{r2}+k_e)t} \quad,
\end{equation}
where
\begin{align}
    f_{r1}&=(1-p_e)p_{r1}/{\rm norm}\\
    f_{r2}&=(1-p_e)(1-p_{r1})/{\rm norm} \notag\\
    f_{e1}&=p_e p_{r1} k_{r1}/(k_{r1}+k_e)/{\rm norm} \notag\\
    f_{e2}&=p_e (1-p_{r1}) k_{r2}/(k_{r2}+k_e)/{\rm norm} \notag\\
    {\rm norm}&=(1-p_e)p_{r1}+(1-p_e)(1-p_{r1})+p_e p_{r1} k_{r1}/(k_{r1}+k_e)+p_e (1-p_{r1}) k_{r2}/(k_{r2}+k_e) \notag \quad.
\end{align}
The probability $p_{r1}$ of a donor fluorophore to be in the conformation 1 was identified with the median $f_1$ of cells expressing mCerulean only. Similarly, $k_{r1}$ (respectively $k_{r2}$) was identified with the inverse of the median $\tau_1$ (respectively $\tau_2$) of cells expressing mCerulean only. The total decay function was then convolved with the IRF, normalized, and fitted to the experimental decay. The remaining fitting parameters, $p_e$ and $1/k_e$, as well as the background intensity and the IRF time shift, were derived by maximum likelihood estimation. This fitting routine is described in Fig.~\ref{fig:FRET}f. The script we used for analysis of FLIM-FRET data is publicly available under \url{https://gitlab.com/polffgroup/actincortex_flim-fret_fitting}.

\subsection{Fitting of model parameters}
\label{sec:SecModelFit}
According to our previously introduced model of binding kinetics \cite{hosseini2020}, the defining equations of homodimer dynamics at a uniform cortex are
\begin{eqnarray}
\partial_t c_{cyt}( t) &=& - 2 k_{on}\frac{S_{cell}}{V_{cell}} c_{cyt}(t) + \frac{\koffone }{V_{cell}}\int_S c_{sb}({\bf x}, t), \label{eq:CrossLinkDyn}\\ 
\partial_t c_{sb}({\bf x}, t) &=& -k_{on,2} c_{sb}({\bf x}, t) +2  \koffzw c_{cl}({\bf x}, t)+2 k_{on}c_{cyt}(t) -\koffone  c_{sb} ({\bf x}, t) \notag\\
\partial_t c_{cl}({\bf x}, t) &=& k_{on,2} c_{sb}({\bf x}, t) - 2 \koffzw c_{cl}({\bf x}, t) \quad, \notag
\end{eqnarray}
where  
$c_{cyt}( t) $ is  a well-stirred cytoplasmic population of cross-linker proteins, 
$c_{sb}(t)$ is the concentration of cross-linker proteins bound at the cortex with a single actin-binding-domain (experiencing no mechanical loading) and $c_{cl}({\bf x}, t)$ is the concentration of cross-linking proteins. 
Here $S_{cell}$ and $V_{cell}$  are the cell surface area and the cell volume, respectively, and $\int_S c_{sb}$  denotes the integration of $c_{sb}$ over the cell surface. Furthermore, $\sigma$ denotes the cortical tension of the cell.  As before \cite{hosseini2020}, we assume a simple catch bond behavior of the cross-linker with a  linear tension-dependence of cross-linking lifetime 
\begin{equation}
\koffzw^{-1} =(1 + \alpha\,\sigma_{})\koffone^{-1}. \label{eq:TaylorExpRate}
\end{equation}
We note that the fraction of cross-linking molecules at the cortex $f_{cl}=N_{cl}/(N_{cl}+N_{sb})$ is given by the formula
\begin{equation} 
    f_{cl}=\frac{k_{on,2}}{k_{on,2}+2 \koffzw} \quad, \label{eq:fcl}
\end{equation}
where $N_{cl}$ and $N_{sb}$ denote the overall number of cross-linking and singly bound molecules at the cortex, respectively. 

Assuming that the fraction of bleached protein is small allows to anticipate that the cytoplasmic concentration is roughly unchanged through the process. In this way, an analytical solution for the recovering cortical concentration fields $c_{sb}(t)$ and $c_{cl}(t)$ can be obtained. In particular, the total cortical  concentration of the recovering population of cross-linkers is
 \begin{eqnarray}
 \left(c_{sb}(t)+c_{cl}(t)\right)&=&  \left(c_{sb}(0)+c_{cl}(0)\right)+ \label{eq:FRAP}\\
&& \hspace{-3cm}
\frac{k_{on}N_{tot} \left(\left[k_{off}(0)(k_{on,2}-2 k_{off}) +(2 k_{off}+k_{on,2})^2\right]
 \left(e^{-\kdetpr t}-e^{-\kdet t}\right)
 -\tilde{k}(2 k_{off}+k_{on,2}) \left(e^{-\kdet t}+e^{-\kdet^\prime t}-2\right)\right)}{2  \tilde{k} S_{cell} \left(\frac{k_{off}(0)k_{off} V_{cell}}{S_{cell}}+k_{on} (2 k_{off}+k_{on,2})\right)} \notag  \quad.
 \end{eqnarray}
 This formula can be used for a fitting of the measured recovery curves of cortical fluorescence intensity from FRAP experiments. %
Here, $S_{cell}$ and $V_{cell}$ are the surface area and the volume of the monitored cell, respectively. The rate constants $k_{det}, \kdetpr$ and $\tilde{k}$ are
\begin{eqnarray}
k_{det}(\sigma_{})&=&\frac{1}{2} \left(\koffone +2 \koffzw +k_{on,2}-\sqrt{-8 \koffone  \koffzw +\left(\koffone +2 \koffzw +k_{on,2}\right)^2}\right) \label{eq:kdet}\\
k_{det}^\prime(\sigma_{})&=&\frac{1}{2} \left(\koffone +2 \koffzw +k_{on,2}+\sqrt{-8 \koffone  \koffzw +\left(\koffone +2 \koffzw +k_{on,2}\right)^2}\right) \label{eq:kdetprime}\\
\tilde k(\sigma_{})&=&\sqrt{-8 \koffone  \koffzw +\left(\koffone +2 \koffzw +k_{on,2}\right)^2} \quad.
\end{eqnarray}

Using the same kinetic Eq.~\eqref{eq:CrossLinkDyn}, we can further simulate the time dynamics of the cortex-to-cytoplasm concentration ratio upon tension peaks. To this end, we used a simple Euler-forward integration method, see matlab script `FRAPandSquishfitting\_Publ.m' in  \url{https://gitlab.com/polffgroup/actin_cortex_modelparameter_fitting}.
In particular, this simulation took into account the increase of f-actin following imposed cell deformation which we found to be well-captured by an exponential relaxation with timescale $\tau_{act}=28\,$s, see Fig.~\ref{fig:Squish}b. 
As f-actin is the substrate for cross-linker binding, a proportional change in cross-linker binding rate over time was assumed.
More concretely, the attachment rate of the cytoplasmic population was assumed to adopt time-dependent values
\begin{eqnarray}
k_{on}(t)=k_{on} S_{cell}^{bef}/S_{cell}(t)\left(1+\beta\,\Theta(t)\left[1-\exp(-t/\tau_{act})\right]\right) \quad,
\label{eq:kont}
\end{eqnarray}
where $\beta=(S_{cell}^{end}/S_{cell}^{bef}-1)$ is the relative surface increase through cell squeezing and $S_{cell}^{end}$, $S_{cell}^{bef}$ and $S_{cell}(t)$ denote the total cell surface area before deformation, at the end of the deformation step and during the deformation time course, respectively. $\Theta(t)$ denotes the Heaviside step function. Note, that the first prefactor $S_{cell}^{bef}/S_{cell}(t)$ in Eq.~\eqref{eq:kont} captures the decrease in binding rate due to dilution of the f-actin substrate during the deformation-induced fast surface area increase. The term proportional to $\Theta(t)$ in Eq.~\eqref{eq:kont} captures the increase of the attachment rate due to the gradual f-actin increase at the cortex after squeezing progress has stopped.

We determined the model parameters, by fitting cell-averaged FRAP curves and the cell-averaged squeezing curves at the same time with a least squares fit. Thereby, we obtained model fits as presented in Table \ref{tab:STab1} and Fig.~\ref{fig:Squish}d-g (fits are dashed black lines).
The mutual weighting of the two parts of the data (FRAP curves and squeezing curves) were adjusted such that both datasets were well captured by the respective fitted curves.

\subsection{Bootstrapping for uncertainty analysis of model parameters}
\label{sec:SecBootstrap}
In brief, model parameters of cross-linker binding dynamics $k_{on}, k_{on,2}, k_{off}(0)$ and $\alpha$  were determined via joint fitting of cell-averaged FRAP recoveries and cell-averages of the time evolution of cortex-to-cytoplasm ratios of fluorescence intensity.
To determine uncertainties of obtained parameters, we performed a boot-strapping approach where subsets of FRAP and squeezing data were generated by omission of one FRAP recovery curve and squeezing data of one measured cell at a time. 
For each data subset, respective FRAP and squeezing data averages were calculated and averaged curves were fitted by the model as described in the preceding section.  
The list of obtained model parameters corresponding to the generated group of data subsets was 100 or larger. 
The corresponding statistics of parameter values was used to obtain error estimates for model parameters, see Fig.~\ref{fig:ViolinPlots}. 

The matlab code used to estimate uncertainties of fit parameters via bootstrapping is publicly available on \url{https://gitlab.com/polffgroup/actincortexanalysis_bootstrapping}.

\subsection{Theoretical model of two-fold mechanosensitivity}
In Fig.~\ref{fig:BlueLight_1}h, we show a plot that illustrates how the linear combination of two different kinds of mechanosensitive effects can account for the observed concentration evolutions of cross-linkers after blebbistatin photoinactivation. The used functions for the direct and indirect effect were 
$f_{\rm dir}(t)=(1-\exp(-t/\tau_1))\exp(-t/\tau_2)$ 
and
$f_{\rm ind}(t)=(1-\exp(-t/\tau_3))$, 
respectively. Chosen parameters were $\tau_1=100$~s and $\tau_2=\tau_3=200$~s. The combined effect was a linear combination $A_{\rm dir} f_{dir}(t)+A_{\rm ind} f_{\rm ind}(t)$ with $A_{\rm dir}=1.5$ and $A_{\rm ind}=0.4$.

\section*{Author Contributions}
V. R. performed the experiments and analyzed the data. V. R. and E. F.-F. designed the experiments. A. H. and M. S. contributed to the planning and data analysis of FLIM-FRET experiments. V.R. and E. F.-F. wrote the manuscript.

\section*{Conflict of interest}
There are no conflicts to declare.

\section*{Data availability}
The data supporting the findings of this study and material are available on request from the corresponding author E.F.F.  

\section*{Code availability}
The modelling and fitting procedures  are described in the Materials and Methods Section and publicly available on GitLab.

\section*{Acknowledgments}
EFF acknowledges financial support from the Deutsche Forschungsgemeinschaft under Germany's Excellence Strategy, EXC-2068-390729961, Cluster of Excellence Physics of Life of TU Dresden.
Furthermore, EFF was funded by the Deutsche Forschungsgemeinschaft (DFG, German Research Foundation) – project number 495224622 (FI 2260/8-1) and by the grant FI 2260/4-1.
MS acknowledges financial support by BMBF (OptiZeD 03Z22E511) and core funding by TU Dresden. 
In addition, the authors thank the CMCB and PoL Light Microscopy Facility for excellent support. We further thank Ina Poser and Anthony Hyman for the provision of the transgenic HeLa cell lines expressing FLNA-GFP, FLNB-GFP and MYH9-mKATE2. Finally, we thank Sean Warren and colleagues who developed the FLIMfit software tool at Imperial College London and made it freely available.

\section*{bibliography}
\bibliographystyle{biophysj}
\bibliography{main.bib}

\vfill
\clearpage
\pagebreak

\begin{appendix}
\setcounter{figure}{0}    
\section{Supplementary material}
\begin{suppenv}
\begin{table}[H]
\centering
\begin{tabular}{|l||c|c|c|c|c|c|}
    \cline{2-7}
     \multicolumn{1}{c|}{}& $k_{on}$ & $k_{on,2}$ & $\koffone$ & $\alpha$ & $f_{cl}$ & $f_{cl}$\\
     \multicolumn{1}{c|}{}& $\mathrm{(\mu m/s)}$ & $\mathrm{(1/s)}$ & $\mathrm{(1/s)}$ & $\mathrm{(10^3m/N)}$ & Ctrl & tension-reduced \\
    \hline\cline{2-7}
     ACTN1 & 9.92e-2 & 1.50e-1 & 3.70e-1 & 5.12e-1 & 34~\% & 20~\%\\
     \hline
     ACTN4 & 5.95e-2 & 4.03e-2 & 1.49e-1 & 1.24 & 35~\% & 17~\%\\
     \hline
     FLNA & 6.81e-2 & 1.92e-2 & 1.55e-1 & 12.4 & 59~\% & 24~\%\\
     \hline
     FLNB & 2.23e-2 & 3.28e-3 & 6.76e-2 & 16.8 & 48~\% & 18~\%\\
     \hline
     MYH9 & 1.25e-2 & 3.34e-2 & 3.74e-2 & 1.19e-1 & 35~\% & 32~\%\\
     \hline
\end{tabular}
\caption{Model parameters $k_{on}, k_{on,2}, \koffone$ and $\alpha$ obtained by fitting of averaged FRAP and cell-squeezing data.
Further, corresponding fractions $f_{cl}=N_{cl}/(N_{cl}+N_{sb})$ (see Eq.~\eqref{eq:fcl}) of cross-linking molecules at the cortex are given for control and tension-reduced conditions using the median values of cortical tension measured during FRAP experiments, see Fig.\ref{fig:FRAP}.
For $\alpha$-actinin-4, these tensions were measured by Hosseini {\it et al.} and were $2.323$ and $0.433~\mathrm{mN/m}$, respectively \cite{hosseini2020}.}
\label{tab:STab1}
\end{table}

\begin{figure}[H]
\centering
\includegraphics[width=0.9\linewidth]{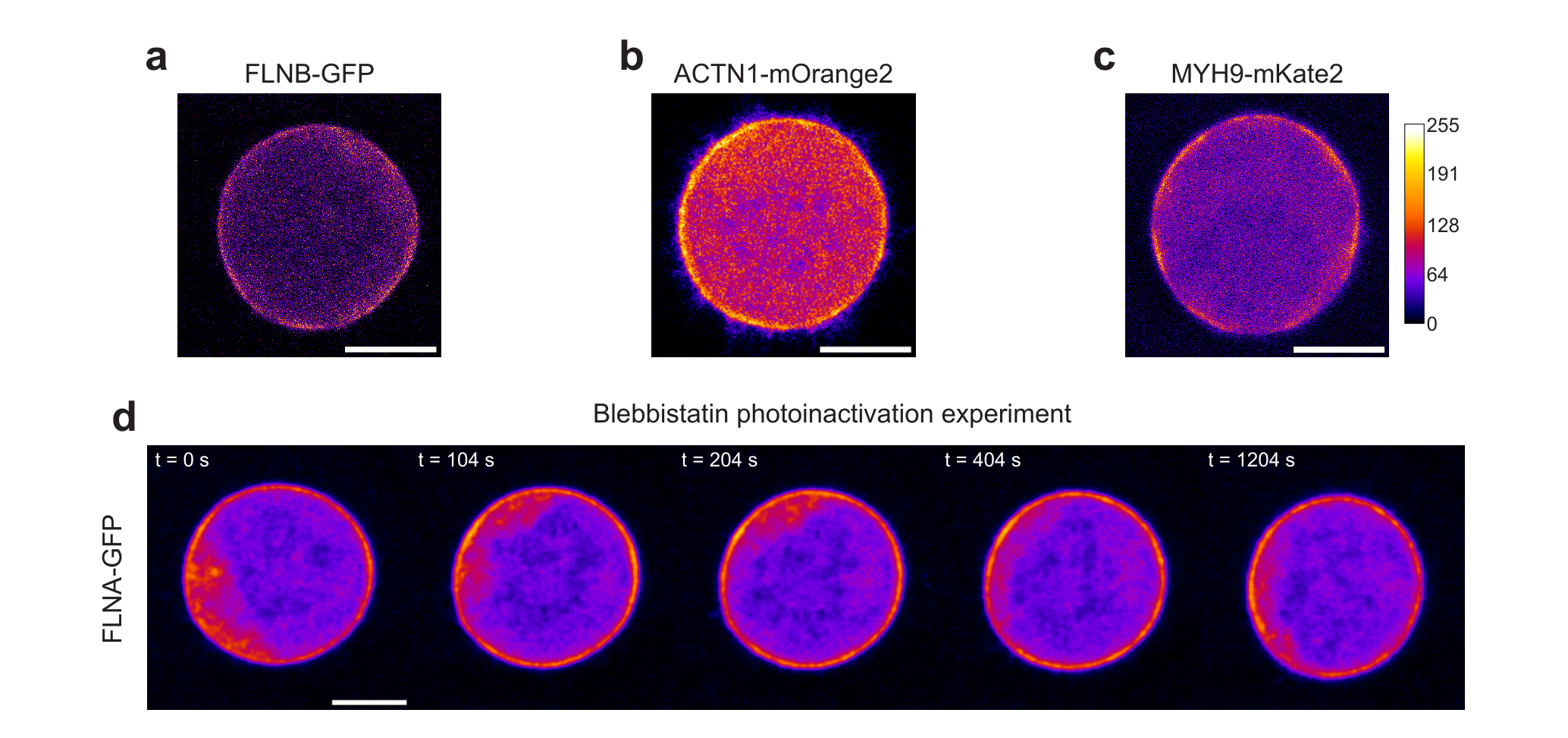}
\caption{\label{fig:S_FRAP_img}
Confocal images of the equatorial cross-section of confined mitotic HeLa cells expressing fluorescently labeled actin cross-linkers.
(a-c) Images acquired before FRAP measurements on cells expressing fluorescently labeled (a) filamin B, (b) $\alpha$-actinin-1 and (c) myosin II in the control condition.
(d) Time series of confocal images acquired during an inhibitor-photoinactivation experiment on a cell expressing fluorescently labeled filamin A. Time $t=0\,$s corresponds to the time point when photoinactivation was started through light exposure. For panel d, a Gaussian blur filter of radius $0.2~\mu$m was applied to make the time dynamics more visible.
Scale bars: 10 µm. The color scale of pixel values is common to all panels. The range of pixel values was adjusted to make the cross-linker localization more visible.
}
\end{figure}

\begin{figure}[H]
\centering
\includegraphics[width=0.9\linewidth]{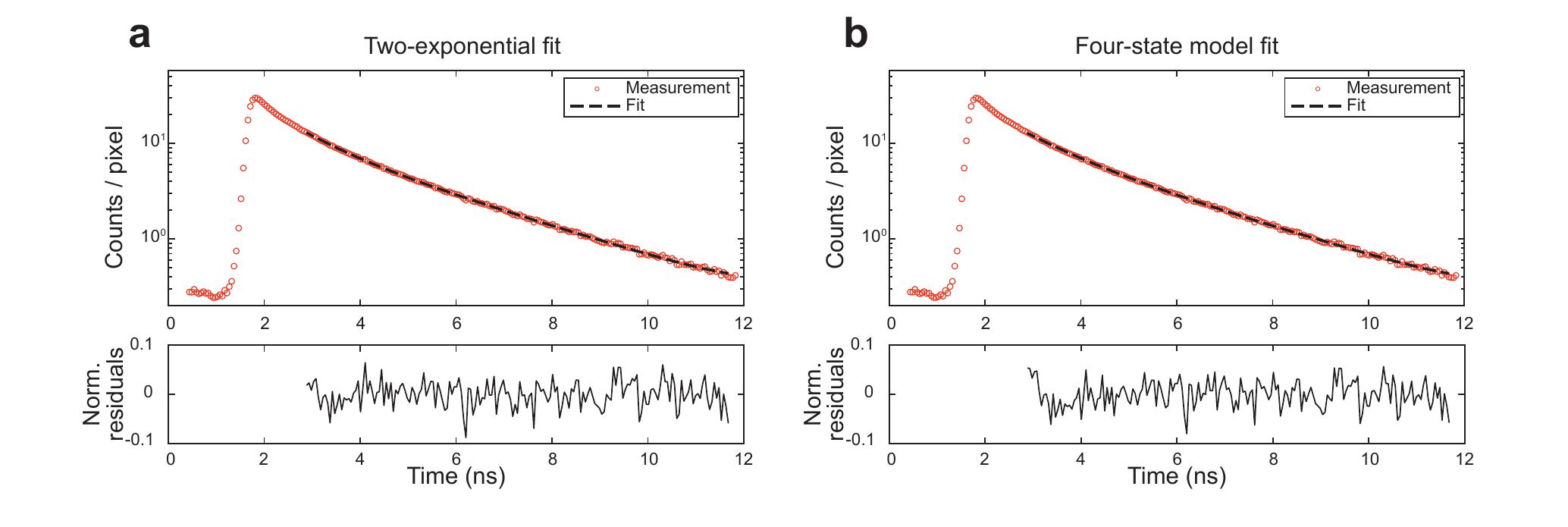}
\caption{\label{fig:S_FLIM-FRET_fit}
Fits of a FLIM decay curve acquired on a mitotic HeLa cell expressing the force-sensing filamin A FRET construct; (a) with the two-exponential decay model and (b) with the four-state fluorescence lifetime model using the analysis procedure illustrated in Fig.~\ref{fig:FRET}f. The blue circles in the top plots represent the measured values and the black dashed lines represent the fit. The bottom plot in each panel shows the residuals divided by the square root of the measured value, over time.
}
\end{figure}

\begin{figure}[H]
\centering
\includegraphics[width=0.9\linewidth]{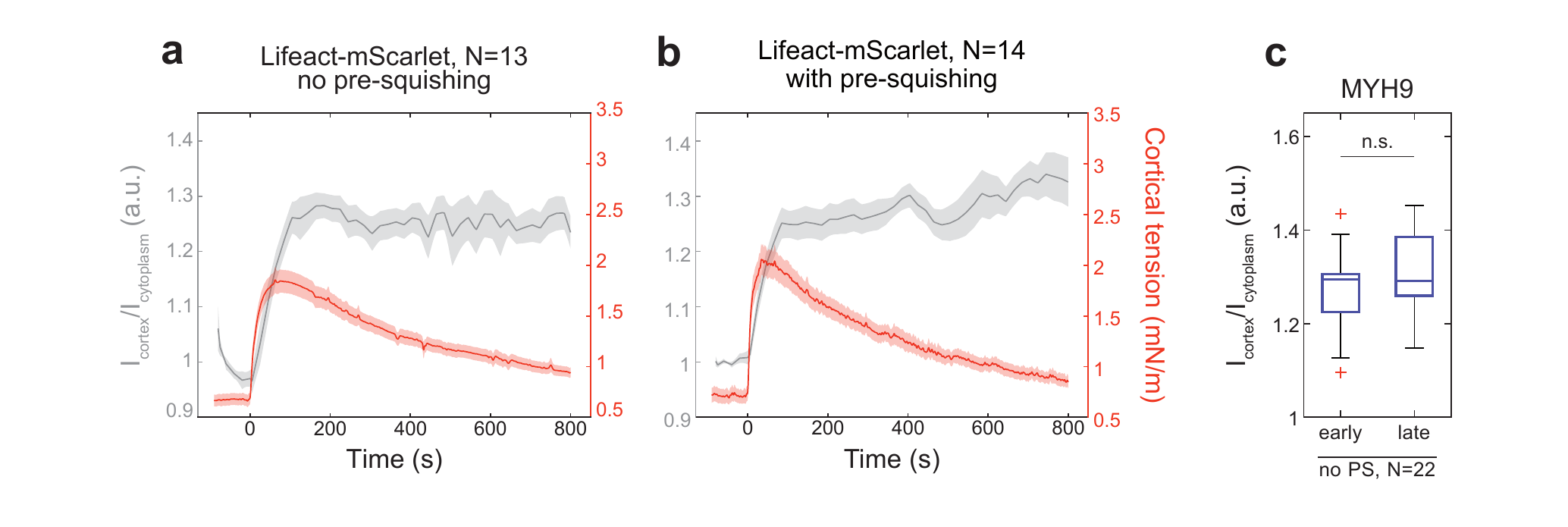}
\caption{\label{fig:S_BlueLight_plots}
F-actin (Lifeact-mScarlet) and myosin II (MYH9-mKate2) intensity increase at the cortex in a step-like manner in response to myosin II inhibitor-photoinactivation.
(a,b) Time evolution of cortical tension and cortex-to-cytoplasm ratio of f-actin (Lifeact-mScarlet ) upon myosin inhibitor-photoinactivation. Here, the cytoplasmic intensity was measured in the outer shell of the cytoplasm (a) without pre-squeezing and (b) with pre-squeezing. The data shown in (a) and (b) correspond to the dashed lines in Fig.~\ref{fig:BlueLight_1}e and Fig.~\ref{fig:BlueLight_2}b, respectively. Shaded grey areas indicate the standard error of the mean.
(c)  Boxplots show the normalized intensity ratios of fluorescently-labeled myosin II measured at the early (150-200 s) and late (750-800 s) time points after inhibitor-photoinactivation without pre-squeezing (see Fig.~\ref{fig:BlueLight_2}a,g). A lack of significant differences between the relative increase at early and late time points indicates that myosin II at the cortex only increases in a short time interval during and right after photoinactivation. 
}
\end{figure}
\end{suppenv}
\end{appendix}

\end{document}